\documentclass[reprint,amssymb,amsmath,aps,superscriptaddress,floatfix,pra,longbibliography]{revtex4-1}

\usepackage{graphicx}
\usepackage{amssymb}
\usepackage{epstopdf}
\usepackage{upgreek}
\usepackage{dcolumn}
\usepackage{bm}
\usepackage{gensymb}
\usepackage{braket}
\usepackage{amsmath}
\usepackage{verbatim}
\usepackage{mathdots}
\usepackage{float}
\usepackage{sidecap}
\usepackage[T1]{fontenc}

\expandafter\ifx\csname package@font\endcsname\relax\else
 \expandafter\expandafter
 \expandafter\usepackage
 \expandafter\expandafter
 \expandafter{\csname package@font\endcsname}%
\fi

\newcommand{\be}{\begin{eqnarray}}
\newcommand{\ee}{\end{eqnarray}}
\newcommand{\bfig}{\begin{figure}}
\newcommand{\efig}{\end{figure}}

\begin{document}

\title{Design of an on-chip superconducting microwave circulator with octave bandwidth}
\author{Benjamin J. Chapman}
\email{benjamin.chapman@colorado.edu}
\affiliation{JILA, National Institute of Standards and Technology and the University of Colorado, Boulder, Colorado 80309, USA}
\affiliation{Department of Physics, University of Colorado, Boulder, Colorado 80309, USA}
\author{Eric I. Rosenthal}
\affiliation{JILA, National Institute of Standards and Technology and the University of Colorado, Boulder, Colorado 80309, USA}
\affiliation{Department of Physics, University of Colorado, Boulder, Colorado 80309, USA}
\author{K.~W. Lehnert}
\affiliation{JILA, National Institute of Standards and Technology and the University of Colorado, Boulder, Colorado 80309, USA}
\affiliation{Department of Physics, University of Colorado, Boulder, Colorado 80309, USA}
\date{\today}

\begin{abstract}
We present a design for a superconducting, on-chip circulator composed of dynamically modulated transfer switches and delays.  Design goals are set for the multiplexed readout of superconducting qubits.  Simulations of the device show that it allows for low-loss circulation (insertion loss $< 0.35$ dB and isolation $>20$ dB) over an instantaneous bandwidth of 2.3 GHz. As the device is estimated to be linear for input powers up to $-65$ dBm, this design improves on the bandwidth and power-handling of previous superconducting circulators~\cite{sliwa:2015,lecocq:2017,chapman:2017b} by over a factor of $50$, making it ideal for integration with broadband quantum limited amplifiers~\cite{macklin:2015,roy:2015,naaman:2017}. 
\end{abstract}

\maketitle

\section{Introduction}

Sophisticated signal-processing often requires that Lorentz reciprocity---the scattering symmetry of an electromagnetic system under exchange of source and detector---be broken. In particular, directionally-routing propagating electromagnetic modes without adding noise or incurring loss is vital for quantum information processing with superconducting circuits.  

Although Maxwell's equations place no restrictions on the bandwidth over which lossless non-reciprocity can occur, synthesis of low-loss on-chip circulators and gyrators has mostly been limited to narrow-band devices~\cite{sliwa:2015,lecocq:2017,chapman:2017b,abdo:2017} (fractional bandwidth $<1\%$).  Ferrite circulators~\cite{fay:1965} provide octave bandwidths, but 
their large magnets make them difficult to miniaturize or integrate with superconducting circuits.
Given the large number of qubits required for fault-tolerant quantum computation~\cite{fowler:2012}, and the standard use of non-reciprocity in qubit readout, circulator performance metrics such as loss, added noise, bandwidth, dynamic range, and size must be simultaneously optimized.  

To address this optimization problem, we propose the superconducting implementation of a recent concept for broadband circulation based on active modulation and delay~\cite{galland:2013,yang:2014,chapman:2017c,lu:2018a,biedka:2018}. 
The approach requires that the delay be commensurate with the modulation frequency, but imposes no constraints on the frequency of the input signal.  This aspect is responsible for the concept's bandwidth, allowing for operation all the way down to dc~\cite{biedka:2018}.  It improves on previous designs~\cite{reiskarimian:2016,biedka:2017,dinc:2017,rosenthal:2017} for broadband active circulators by engineering the desired circulation in a hardware efficient manner, which may in principle be lossless. Such schemes are complementary to superconducting travelling-wave isolators, which provide an alternate path to compact broadband non-reciprocity~\cite{ranzani:2017}.  

When modeled as a superconducting integrated circuit, simulations with realistic parameters show our design allows for low-loss circulation ($<0.35$ dB) with octave bandwidth and nW scale power-handling.  With respect to bandwidth and linearity, this represents a 50-fold improvement over other near-lossless superconducting circulators~\cite{sliwa:2015,lecocq:2017,chapman:2017b}, well-suited for integration with broadband quantum-limited amplifiers~\cite{macklin:2015,roy:2015,naaman:2017}.

\section{Theory of operation}
\label{sec:theory}
\begin{figure*}[hbt]
\begin{center}
\includegraphics[width=2.0\columnwidth]{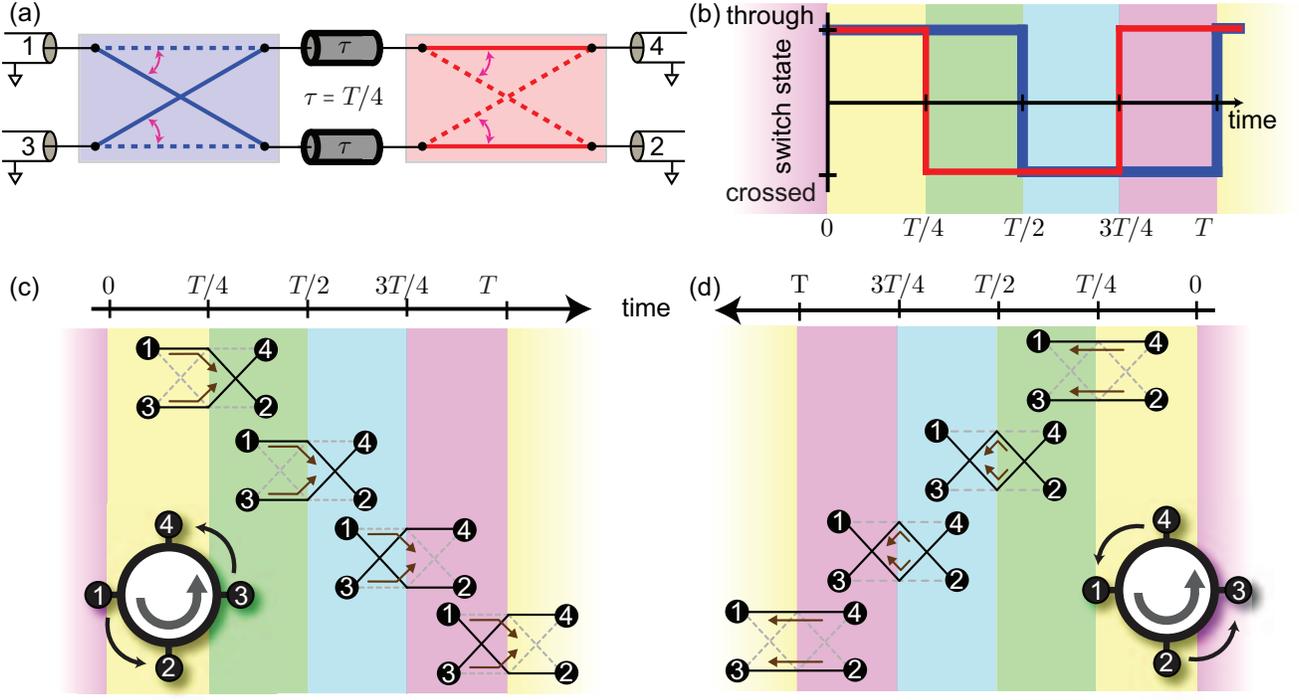}
\caption {\textbf{Circulation with actively modulated transfer switches and delays.} (a) A four-port circulator is constructed from two dynamically modulated transfer switches and a pair of delay lines. The left (blue) switch is shown in the crossed state, and the right (red) switch is shown in the through.  The delay is set to one quarter of the modulation period $\tau = T/4$. (b) The left and right switches are modulated with period $T$ and a relative phase of $\pi/2$. For visual clarity, the modulation period is divided into quarter-period intervals indicated by the pastel background colors. (c) The schematics depict the way the switches will route a right-propagating signal, depending on its arrival time.  Signals incident on port 1 are always routed to port 2, and signals incident on port 3 are routed to port 4.  Note that these are not ``snap-shots'' of the switches' configurations, as the signal is delayed by a quarter-period of the modulation as it propagates through the network.  (d) The same depiction as (c), but for left-travelling waves incident on ports 2 or 4. To keep the device's orientation the same as in (b), time flows from right to left in this diagram.
}
\label{fig:fig1}
\end{center}
\end{figure*}

The proposed circulator is composed of two transfer switches connected by a pair of delay lines (Fig.~\ref{fig:fig1}a).  The transfer switches have four ports, and are dynamically modulated with period $T$, toggling between their ``through'' and ``crossed'' states. Figure~\ref{fig:fig1}b shows how the two switches are periodically actuated by square waves with a relative phase of $\theta = \pi/2$. The delays are chosen to have duration $\tau = T/4$.

To see how the circulation is created, consider how a right-propagating wave packet (e.g. a signal-burst incident on ports 1 or 3) will be scattered by the network.  Figure~\ref{fig:fig1}c shows how the two switches will route the incident signal, depending on the portion of the modulation period in which it arrives.  For example, if the signal arrives in the first quarter-period of the modulation (the yellow time-window), the left (blue) transfer switch will be in its through state.  But by the time the signal exits the delay line, the right (red) transfer switch will be in its crossed state.  Signals incident on port 1 are therefore routed to port 2, and signals incident on port 3 are routed to port 4.  
This process may be continued for the entire modulation period. One observes that the port-to-port signal-routing is independent of the incident signal's arrival time, despite the fact that the signal may travel along different paths before exiting the network.

The non-reciprocal character of the network may be verified by repeating this analysis for left-travelling wave packets (e.g. signals incident on ports 2 or 4), as depicted in Fig.~\ref{fig:fig1}d.  Again, the port-to-port signal-routing is independent of the signal's arrival time: signals incident on port 2 are always routed to port 3, and signals incident on port 4 are always routed to port 1.  The scattering of the network is not invariant upon exchange of source and detector. It forms a four-port circulator.

This graphical time-domain representation of the circulator may be complemented by a quantitative frequency-domain analysis in which all $m$ sidebands (detuned by integer multiples of the modulation frequency $\Omega = 2 \pi/T$)  
are treated as ports of the network. 
The scattering of a signal incident on the first physical port may then be expressed as
\begin{eqnarray}
\label{circulatorS}
S_{11}^m &=& 0, \\ \nonumber
S_{21}^m &=& \frac{1}{2}(\delta_{m,0} -r_m), \\ \nonumber
S_{31}^m &=& 0, \\ \nonumber
S_{41}^m &=& \frac{1}{2}(\delta_{m,0} +r_m), \\ \nonumber
r_m &\equiv&  -\frac{4}{\pi^2} \sum_k  \frac{e^{j k \Omega \tau} e^{j(m-k)\theta}}{(m-k)k},
\end{eqnarray}
where the sum on $k$ runs over all odd integers, $\delta$ is the Kronecker-delta, and $m$ is an index denoting effective ports at different frequencies, which is taken to be zero at the frequency of the input signal (e.g. $m = -2$ denotes the scattering of signals detuned from the input by $-2\Omega$).  The term $r_m$ describes the transmission of an odd-excitation of the network's first and third ports: its interference with the transmission of an even excitation creates the desired circulation.  
Such a representation is useful in quantifying the effect of imperfections and in evaluating the device's linearity.
(Absent nonidealities, Eq.~(\ref{circulatorS}) yields perfect four-port circulation when $\Omega \tau = \pi/2$, $\theta = \pm \pi/2$, for which $r_m = \pm\delta_{m,0}$).
Appendix~\ref{sec:derivation} includes the full multi-frequency scattering matrix derivation, and extension of Eq.~(\ref{circulatorS}) to include the effects of group delay dispersion, finite modulation bandwidth, and loss.

The device's sensitivity to these non-idealities depends on the modulation rate $\Omega$: Equation~(\ref{circulatorS}) shows 
that group delay dispersion becomes problematic an $\Omega/(2\pi)$ shift in frequency causes a delay variation on the scale of $\tau$. Similarly, 
a finite bandwidth $\Omega_b$ of the square-wave modulation truncates the sum in Eq.~(\ref{circulatorS}) at $k_\textrm{max} = \Omega_b/\Omega$.
Both of these considerations prejudice the design in favor of longer delays and modulation rates which are much smaller than the frequencies of the circulated signals.  Slow modulation also facilitates filtering, allowing suppression of spurious coupling between the modulation and signal lines.

The benefits of slow modulation, however, do not come without a price. Slow modulation is only possible with long delays, which in the context of integrated circuits typically entails greater loss and circuit size and/or reduced bandwidth.  
In a superconducting design space, though, vanishing conductive losses make long, low-loss, broadband delays possible
~\cite{hohenwarter:1993,wang:2005,su:2008,zhong:2018}, and circuit size is the only remaining penalty. We leverage this fact in the proposed superconducting implementation of a broadband, low-loss, on-chip circulator, the topic of the following section.

\section{Proposed superconducting implementation}
We set design goals for our circulator by anticipating its use in superconducting qubit readout.  In particular, we design for integrability with broadband near-quantum-limited amplifiers such as Josephson parametric amplifiers~\cite{roy:2015} or Josephson traveling-wave parametric amplifiers~\cite{macklin:2015}, in the frequency range of 4 to 8 GHz. 
To allow for robust, wafer-scale production, we work in the design space allowed by optical lithographic fabrication.


The modular nature of the proposed circulator reduces the challenge of designing a low-loss, broadband circulator to the problem of designing low-loss, broadband transfer switches and delays.  We address these in turn, and then present simulations of the full device.


\subsection{Transfer switch}
\begin{figure}
\begin{center}
\includegraphics[width=1.0\columnwidth]{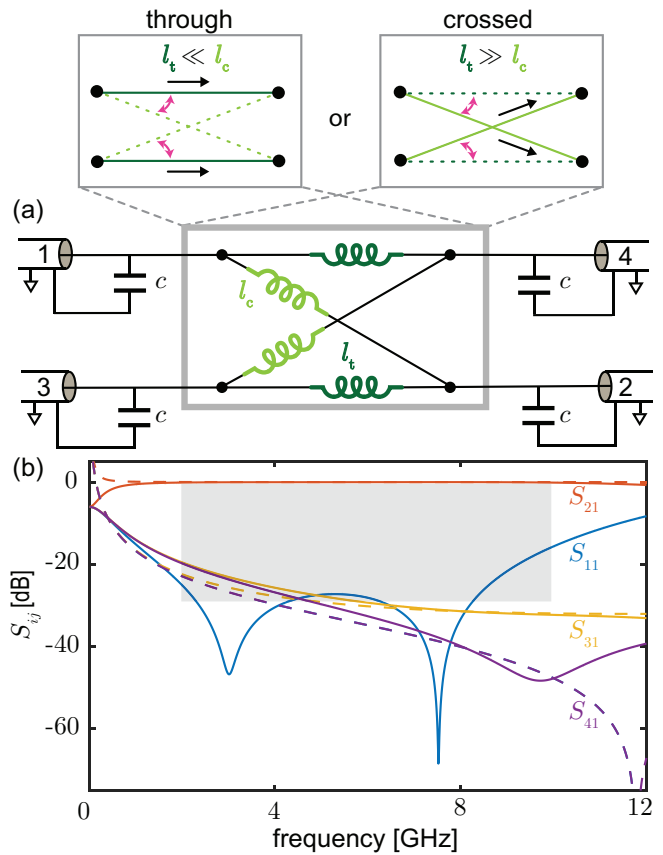}
\caption {\textbf{Transfer switch with tunable inductors.} A symmetric lattice of tunable inductors forms a transfer switch.  (a) When the inductors are tuned such that $l_t \ll l_c$, the switch is in its through state.  Tuning $l_c \ll l_t$ realizes the switch's crossed state. Parallel capacitances $c$ are used to match the switch to a characteristic impedance $Z_0$.  (b) Calculated scattering parameters of the network in (a) parametrized by Eq.~(\ref{epsilon}), with $l_0 = 0.94$ nH, $c = 270$ fF, $\epsilon = 2.5 \times 10^{-2}$.  Solid lines are exact solutions. Dashed lines are the approximations to first-order in $\epsilon$ given in Eq.~(\ref{epsexpansion}). Insertion loss is $<0.03$ dB  between 4 and 8 GHz.  
Gray rectangle shows the Bode-Fano criterion (Eq.~(\ref{BF})) between 2 and 10 GHz. Its proximity to $S_{11}$ attests to the efficacy of the single-pole matching network.  
}
\label{fig:symlattice}
\end{center}
\end{figure}

To construct a fast superconducting transfer switch, we arrange two pairs of nominally identical and tunable inductors in a symmetric lattice (bridge) topology (see Fig.~\ref{fig:symlattice}a). The pairs of inductors tune in a coordinated fashion, allowing the switch to toggle between its through and crossed states (inset of Fig.~\ref{fig:symlattice}a). 

Constructing the switch with purely reactive elements allows it to be completely free of dissipation. Limiting insertion loss
, however, requires impedance matching.  
Synthesis of a matching network is simplified if the bridge can be strongly imbalanced---that is, if the impedance of the through elements may be made much larger than the impedance of the crossed elements, or vice versa.  In this limit, the Bode-Fano criterion~\cite{bode:1945,fano:1950} for the switch is approximately that of a series $RL$ circuit~\cite{pozar:2011}:
\begin{eqnarray}
   -\int_0^\infty \ln{|\Gamma(\omega)|} d\omega < \pi \frac{R}{L}.
   \label{BF}
\end{eqnarray}
Here $R$ is taken to be the characteristic impedance of the surrounding lines $Z_0 = 50$ Ohms, $\Gamma(\omega)$ is the reflection coefficient, and $L$ is the small inductance in the network ($l_t$ in the switch's through state, $l_c$ in the crossed state).  A broadband match may then be obtained for sufficiently small $L$.  For example, limiting reflections to $\Gamma = -20$ dB over an 8 GHz band requires that $L$ be less than approximately 1 nH.  We make this match in a simple, symmetrical way, with shunting capacitors $c$ sandwiching the inductor $L$ (Fig.~\ref{fig:symlattice}a).

To quantify the needed imbalance, we analyze the switch in its ``through'' position and parametrize the pairs of tunable inductors as
\begin{eqnarray}
   l_{t} &=& l_0, \nonumber \\
   l_{c} &=& \frac{l_0}{\epsilon}.
   \label{epsilon}
\end{eqnarray}
Here $\epsilon$ is a small parameter that describes the degree to which the bridge can be imbalanced.  A standard circuit analysis (see Appendix~\ref{sec:switchS}) allows expansion of the switch's scattering parameters in powers of $\epsilon$.  We find that the amplitude of both reflection and unwanted transmission (e.g. transmission to isolated ports of the switch) scale to leading order as $\epsilon$, provided the impedance of the resonator $Z_r = \sqrt{\frac{l_0}{2c}}$ is equal to $Z_0$, and the 
capacitive corner-frequency $f_c \equiv (Z_0 c)^{-1}$ is much larger than the maximum frequency of operation ($\approx10$ GHz).  When $l_0 \approx 1$ nH, this may be accomplished with $c \approx 200$ fF.  Fine tuning of the inductance and capacitance may then be used to adjust the passband of the switch around the center frequency $\omega_0 = 1/\sqrt{2 l_0 c}$.  Made in this way, the imbalanced switch is effectively an RLC oscillator with quality factor of order one.

As the switch's isolation and return loss depend directly on $\epsilon$, implementing inductors which may be tuned over a wide range is a critical task.  
We suggest two ways in which small $\epsilon$ values may be realized.  

The first uses laddered arrays of Josephson junctions which are flux-biased close to frustration.  Devices made in this way have demonstrated $\epsilon \leq 2\times 10^{-2}$~\cite{bell:2012,naaman:2016}.  
The second uses dc-SQUIDs with small, high-aspect ratio loops ($\simeq5$ $\mu$m $\times$ $\simeq100$ nm).  These may be fabricated with optical lithography by orienting the loop such that its normal vector is parallel to the chip's surface.  We expect $\epsilon<2.5\times10^{-2}$ to be possible with such a design. (For more details, see Appendix~\ref{sec:SQUIDs}). 
Implemented in either fashion, the tunable inductors may then be arranged in a bridge geometry and tuned with a pair of flux controls, as in Refs.~\cite{kerckhoff:2015,chapman:2016,chapman:2017,chapman:2017b}.

Figure~\ref{fig:symlattice}b shows the magnitudes of the calculated scattering parameters of our switch design, optimized for the 4 to 8 GHz band given $\epsilon = 2.5 \times 10^{-2}$.  Between 4 and 8 GHz, insertion loss is $<0.03$ dB, isolation exceeds $25$ dB, and reflections are less than $-26$ dB.  The impedance match is therefore comparable to the Bode-Fano limit for an $RL$ circuit (shaded gray rectangle).  The switch's group delay dispersion is non-zero, but as the total delay provided by the switch is approximately the inverse of its bandwidth (and $Q \approx 1$), it imparts a delay which is small relative to that of the delay lines.

Imbalancing an inductive bridge to this degree puts constraints on the junction's plasma (self-resonance) frequency $\omega_p$. When the tunable inductors are flux-biased into their high-inductance state, their Josephson inductance changes by a factor of $1/\epsilon$, decreasing the plasma frequency of the unbiased junction by a factor of $\sqrt{\epsilon}$.  To ensure that this reduced plasma frequency is always greater than the circulator's operation band, we require $\omega_p > 2 \pi \times \sqrt{\epsilon} \times 10$ GHz in the unbiased junctions.  
For a niobium trilayer process where $\omega_p$ is a known function of the junction critical current density $J_c$~\cite{maezawa:1995,tolpygo:2017}, the above constraint defines a minimum $J_c$.  Given a minimum feature size of $2~\mu$m, the minimum junction critical current is then set at $I_0 = 9$ $\mu$A.  Finally, the target array inductance of $l_0 \approx 1$ nH can be achieved by cascading $N \approx 40$ of the SQUIDs in series.


An auxiliary benefit of this large plasma frequency requirement comes in the device's linearity.  Similar broadband switches fabricated with $35$ $\mu$A junctions have reported $1$-dB compression points at $-53$ dBm~\cite{naaman:2016}.  Assuming power-handling scales with the square of junction critical currents, the design proposed here would saturate around $-65$ dBm, ample for integration with Josephson travelling-wave amplifiers~\cite{macklin:2015}.

For use in the proposed active circulator, it is not enough for the transfer switches to perform statically: they must also be fast.  In principal, the only physical constraint on the switching speed is the junction self-resonance frequency.  In practice, however, further restriction of the modulation bandwidth can limit the extent to which photons in the flux control lines couple into the circuit and emerge as noise in the operation band.  

In particular, in the regime where the modulation rate $\Omega$ is much less than the 4 to 8 GHz operating frequencies, it is advantageous to limit the modulation bandwidth to half of the minimum operating frequency.  This provides ample separation between the modulation band ($1/T$ to 2 GHz) and the signal band (4 to 8 GHz), which facilitates filtering.  It also ensures that modulation tones which spuriously couple into the signal lines will be below the signal band, even if they are mixed up in frequency by the second transfer switch.  Further discussion of the control waveforms' effect on the circulator's output spectrum is given in Appendix~\ref{sec:noise}.



\subsection{Delay lines and optimal modulation rate}
Modulating the transfer switches slowly requires a broadband and low-loss delay on the scale of ns, many microwave periods long. Furthermore, this delay must be relatively free of dispersion. These considerations suggest the use of a transmission line delay which supports a TEM or quasi-TEM mode, such as a coplanar waveguide (CPW).  Such a line may be tightly meandered to realize the desired broadband delay in a compact manner. 

The primary sources of dispersion in such a structure are the resonant modes associated with the ground plane and the bends in the path of the CPW. To account for these, we conducted numerical simulations of meandered CPWs using a planar method-of-moments solver.  These show 1 ns delays may be achieved over the entire 2 to 10 GHz band, with return loss better than $30$ dB and fractional group velocity dispersion $<3\times10^{-4}$ per $80$ MHz. Dispersion at this level would cause $0.05$ dB of circulator insertion loss (see Appendix~\ref{sec:derivation}).  (Dispersion from the quasi-TEM nature of the CPW mode, and the frequency-dependence of niobium's surface inductance, are irrelevant on this scale~\cite{hasnain:1986,gao:2008}.)


At temperatures and frequencies where the thermal and photon energies are much less than the superconducting gap, dissipation in a CPW is dominated by dielectric loss.  For a signal with frequency $\omega$ propagating in a TEM mode for a time $\tau= \pi/(2\Omega)$, the attenuation $a$ from the surrounding dielectric is~\cite{orfanidis:2002}
\begin{eqnarray}
   a = \textrm{exp}\left(-\frac{\omega \tau}{2} \tan{\delta}\right).
   \label{dielectricloss}
\end{eqnarray}
Here $\tan{\delta}$ is the dielectric loss tangent.  Coplanar waveguides fabricated on silicon substrates routinely demonstrate quality factors $>10^{5}$~\cite{oconnell:2008}, making the dissipation from a $1$ ns delay less than 3 mdB for signals in our band.



To estimate the physical size of such a delay and the resulting circuit, we note that for a CPW on silicon where microwaves travel more slowly than the vacuum speed of light $c_0$, 
the needed delay line has length $d \approx \pi c_0/(5 \Omega)$. 
Assuming a minimum lithographic feature size of $2$~$\mu$m, a 50 Ohm CPW may be as narrow as 10 $\mu$m, and a reasonable value for the meander pitch is $p = 60$~$\mu$m.  
The switches themselves require less than 1 mm$^2$, so the entire layout may fit on a chip with area $2dp$ + 2 mm$^2$.  

\begin{figure}[hbt]
\begin{center}
\includegraphics[width=1.0\columnwidth]{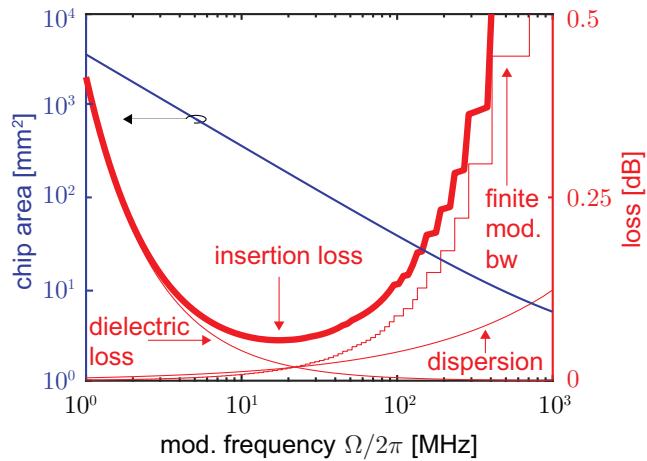}
\caption {\textbf{Optimal modulation rate.} Estimated chip area (left axis, blue trace) and total insertion loss (right axis, thick red trace) as a function of the modulation frequency.  Contributions to the insertion loss from dispersive, dielectric, and finite modulation bandwidth effects are also shown.  These assume a fractional group velocity dispersion of $3\times10^{-4}$ per $80$ MHz, a dielectric loss tangent of $\tan{\delta} = 10^{-5}$, and an $\Omega_b/(2\pi) = 2$ GHz modulation bandwidth. 
}
\label{fig:optimal}
\end{center}
\end{figure}

Selecting a precise value for the modulation frequency $\Omega$ therefore amounts to balancing competing effects: slower modulation minimizes the loss associated with dispersion and finite modulation bandwidth, but increases dielectric losses and circuit size.  Figure~\ref{fig:optimal} summarizes these trade-offs.  The blue trace shows circuit size (left axis), while the red traces shows the various loss channels (right axis), calculated with Eq.~(\ref{dielectricloss}) and the methods described in Appendix~\ref{sec:derivation}. A modulation rate of $\Omega = 2\pi \times 80$ MHz ($d\approx 37.5$ cm) is estimated to keep the circulator's insertion loss below $0.1$ dB while allowing the device to fit on a $7$ mm by $7$ mm chip.

\subsection{Simulated Performance}
To predict the performance of the proposed superconducting circulator we conduct time-domain numerical simulations of the design.  The delays are modeled as ideal transmission lines which provide a delay of $\tau = \pi/(2 \Omega) = 3.125$ ns (commensurate with $\Omega = 2 \pi \times 80$ MHz).  The transfer switches are modeled by the lumped-element network depicted in Fig.~\ref{fig:symlattice}a, with several non-idealities included:  

First, each inductor is replaced by an array of $N = 46$ inductors in series, each of which is assumed to be tunable by an external signal $\Phi_e$, via the relation 
\begin{eqnarray}
   l(\Phi_e) = \frac{l_0}{\sqrt{\epsilon^2 + (1-2\epsilon) \cos^2{\left(\Phi_e\right)}}}+l_g.
   \label{TunableInductance}
\end{eqnarray}
This functional form is chosen to replicate the flux-dependence of an asymmetrical dc-SQUID's inductance~\cite{vanduzer:1981}, and make contact with Eq.~(\ref{epsilon}) in the limits $\Phi_e \to 0$ and $\Phi_e \to \pi/2$.  We fix $N l_0 = 1.02$ nH and $\epsilon = 2.5 \times 10^{-2}$, and the geometric inductance is chosen to be $N l_g = 207$ pH.  Second, alongside each of these tunable inductors we place a parallel capacitance of $c_J = 180$ fF to mimic the self-capacitance of the junctions.  Lastly, the tunable inductors are modulated by two approximately square-wave signals with frequency $\Omega = 2 \pi \times 80$ MHz.  We model the finite modulation bandwidth by synthesizing the bias signals from the sum of the first 25 Fourier components of a square-wave, such that the bias waveform has no spectral weight above $\Omega_b/(2 \pi) = 25 \Omega/(2 \pi) = 2$ GHz.

\begin{figure}
\begin{center}
\includegraphics[width=1.0\columnwidth]{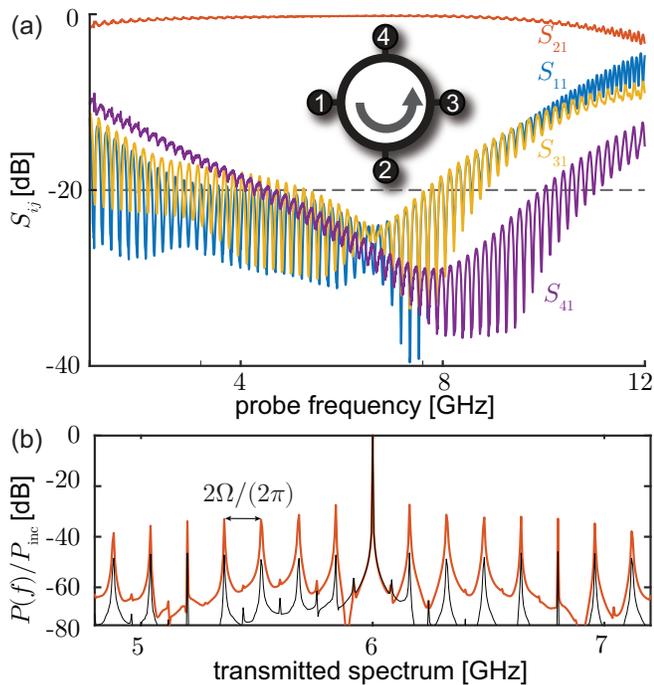}
\caption {\textbf{Simulated performance.} Time-domain numerical simulations of the circulator, with transfer switches formed from symmetric lattices of tunable inductors modulated at $\Omega = 2\pi\times 80$ MHz.  (a) First column of the simulated scattering matrix, when the device is configured for counter-clockwise circulation.  The horizontal line at $-20$ dB is a guide to the eye. (b) Power spectrum of the signal emerging from the circulator's second port, when the circulator is driven by a 6 GHz tone incident on its first port.  Power is normalized to the incident signal power $P_{\textrm{inc}}$.  The thin black trace shows the same quantity in a simulation with infinite modulation bandwidth.}
\label{fig:simulink}
\end{center}
\end{figure}

Figure~\ref{fig:simulink}a shows the first column of the simulated network's scattering parameters.  
Between 5.9 and 8.2 GHz, the circulator's isolation exceeds 20 dB.  Insertion loss in the same band is less than $0.3$ dB.  Accounting for loss caused by disperion in the CPW delays (which is not present in the time-domain simulation), the total loss of the device is expected to be less than $0.35$ dB across this band.

The primary performance limitation is the finite modulation bandwidth: if the simulation is repeated with infinite modulation bandwidth and all other parameters held fixed, isolation is greater than $20$ dB between $4$ and $8.5$ GHz, and insertion loss is improved to better than $0.1$ dB.  Power emerging out of the circulator's other three ports (at the input frequency) accounts for $0.03$ dB of this loss.  We attribute the final $0.05$ dB to the dispersion of the switches.


The simulated $~0.2$ dB of insertion loss that results from an $\Omega_b = 2\pi \times 2$ GHz modulation bandwidth is greater than the $0.07$ dB expected from the analysis in Appendix~\ref{sec:derivation}.
The discrepancy arises from the sensitive dependence of the inductance on the control signal near the $\Phi_e = \pi/2$ operating point (at $\Phi_e = \pi/2$, $dl/d\Phi_e$ diverges).  
While the rising-edge of the finite bandwidth flux signal  approaches (and then overshoots) its target value of $\pi/2$, the efficacy of the switches is diminished until the control flux settles.

The ripple in the scattering parameters has a free spectral range of $2\Omega$, and is caused by small reflections at the second transfer switch.  As these reflections take two extra passes through the delay lines, they experience an additional delay of $2 \tau = \pi/\Omega$ before interfering with the un-reflected signal.  

Despite the coordinated modulation used to create circulation, the majority of the transmitted power is unchanged in frequency.  The orange trace in Fig.~\ref{fig:simulink}b shows the spectral content of the signal exiting the circulator's second port, when the first port is driven by a $6$ GHz tone.  Most of the power ($-0.28$ dB) emerges at 6 GHz.  Sidebands detuned from $6$ GHz by multiples of $2 \Omega/(2\pi)$ are also visible.  The largest of these sidebands is less than $-27$ dB. For reference, the thin black trace also shows the same simulation with infinite modulation bandwidth.  Although they are suppressed by over $45$ dB, modulation sidebands are still present in the spectrum.  They are caused by the finite isolation of the transfer switches.

\section{Conclusion}



We have described the theory of operation for a broadband and low-loss circulator based on active modulation and delay.  The device may be understood as a network of two simple components: transfer switches and delays.  Design of a superconducting implementation was presented and discussed in terms of the optimal modulation rate $\Omega$, a key design parameter.  The device was then simulated numerically: isolation is greater than $20$ dB while insertion loss is less than $0.35$ dB over a bandwidth of $2.3$ GHz.  Based on comparison with  demonstrated Josephson junction based switches, its $1$-dB compression point is expected to be $-65$ dBm. 

The proposed circulator therefore represents a major advance in on-chip non-reciprocity, improving on the bandwidth and power-handling of other near-lossless devices~\cite{sliwa:2015,lecocq:2017,chapman:2017b} by more than a factor of $50$.  These gains become especially powerful when the broadband circulator is combined with a linear and broadband near-quantum limited amplifier~\cite{macklin:2015,naaman:2017}.
Such a quantum microwave receiver, which would be linear at powers $10^6$ times greater than a typical qubit readout tone, and operable over a bandwidth $10^3$ times greater than a typical readout-cavity's coupling rate, could allow for the simultaneous readout of many qubits in an on-chip package. 


\vspace{0.1in}
\noindent{\emph {Acknowledgment}} The authors thank Jiansong Gao, Bradley Moores, and Xian Wu for helpful discussions.  E.I.R acknowledges support from the ARO QuaCGR fellowship.  This work is supported by the ARO under contract W911NF-14-1-0079 and the National Science Foundation under Grant Number 1125844.



\appendix
\section{Circulator scattering parameters}
\label{sec:derivation}
Analysis of the network in Fig.~\ref{fig:fig1}a is simplified by separately considering even and odd excitations of the network's left and right sides. Let the matrix $\mathbf{U}$ describe the unitary transformation between the numbered port basis and the basis of left/right and even/odd excitations:
\begin{eqnarray}
    \mathbf{U} = \frac{1}{\sqrt{2}} \left(\begin{array}{cccc} 
    1 & 0 & 1 & 0 \\  
    0 & 1 & 0 & 1 \\  
    1 & 0 & -1 & 0 \\ 
    0 & 1 & 0 & -1 \\ 
    \end{array}\right).
\end{eqnarray}
We assume that both the switches and delay lines are impedance matched.  We also initially assume lossless, distortion-less delays and infinite switch-modulation bandwidth.  These last three assumptions are relaxed at the conclusion of the derivation.  

Even excitations are unaffected by the state of the switches. As all of the network's components are impedance matched by assumption, common mode excitations are unaltered as they propagate through the network.  The scattering matrix for the even mode excitations $\mathbf{S_e}$ is therefore:
\begin{eqnarray}
    \mathbf{S_e} = \left(\begin{array}{cc} 
    0 & 1 \\  
    1 & 0 \\ 
    \end{array}\right).
    \label{Se}
\end{eqnarray}

To analyze the network's odd dynamics, we introduce a vector whose entries describe the spectral content $a_\omega$ at the sidebands of the modulation frequency $\Omega$:
\begin{eqnarray}
    \mathbf{a_\omega} = \left(\begin{array}{c} 
    \vdots \\
    a_{\omega + 2 \Omega} \\  
    a_{\omega + \Omega} \\
    a_\omega  \\
    a_{\omega - \Omega} \\
    a_{\omega - 2 \Omega} \\
    \vdots \\
    \end{array}\right).
\end{eqnarray}
We choose to index this infinite-dimensional vector with positive and negative indices, and designate the term $a_\omega$ as the 0 index.

Let us define the modulation of a single transfer switch which is square-wave biased with phase $\phi$ as $H(t) \equiv \textrm{sign}\left[ \sin(\Omega t + \phi) \right]$.  By Fourier decomposition,
\begin{eqnarray}
    H(t) &=& \frac{2}{j\pi} \sum_n \frac{e^{jn\phi} e^{jn \Omega t}-e^{-jn\phi} e^{-jn \Omega t}}{n},
\end{eqnarray}
where the sum runs over positive odd integers.

Written in this way, the action of a switch on the vector $\mathbf{a_\omega}$ can be expressed as a matrix. In element-wise form, this matrix may be defined piece-wise as
\begin{eqnarray}
    \mathbf{H}_{mn}(\phi) =    \left\{
\begin{array}{ll}
      0 &\textrm{if $m-n$ even,}\\
      \frac{2e^{j(m-n)\phi}}{j\pi(m-n)} & \textrm{if $m-n$ odd.}\\
\end{array} 
\right.
\end{eqnarray}

The action of the delay lines is represented by a diagonal matrix with elements
\begin{eqnarray}
    \mathbf{D}_{mn} = e^{jm\Omega \tau} \delta_{m,n}.
\end{eqnarray}

We may now calculate the action of the full cascaded network, formed by a left switch with $\phi=0$ (represented by the matrix $\mathbf{L}\equiv \mathbf{H}(0)$), a delay (represented by the matrix $\mathbf{D}$), and a right switch with $\phi = \theta$ (represented by the matrix $\mathbf{R} \equiv \mathbf{H}(\theta)$).  This network transforms a right-propagating differential excitation by the matrix product $\mathbf{R} \mathbf{D} \mathbf{L}$.  If the right-propagating excitation is a pure tone with frequency $\omega$, the output of the network at the $m^{\textrm{th}}$ sideband is
\begin{eqnarray}
    r_m \equiv (\mathbf{R}\mathbf{D}\mathbf{L})_{m0} = -\frac{4}{\pi^2} \sum_k  \frac{e^{j k \Omega \tau} e^{j(m-k)\theta}}{(m-k)k},
\end{eqnarray}
if $m$ is even and $0$ if $m$ is odd. Here the sum on $k$ runs over all (positive and negative) odd integers. The signal transmitted at the input frequency is therefore
\begin{eqnarray}
    (\mathbf{R}\mathbf{D}\mathbf{L})_{00} = \frac{4}{\pi^2} \sum_k  \frac{e^{j k (\Omega \tau-\theta)}}{k^2},
\end{eqnarray}
which is $1$ when $\Omega\tau -\theta = 0$.

For a left-propagating wave, the action of the network is given by the reverse-ordered product, $\mathbf{L} \mathbf{D} \mathbf{R}$. For a pure tone at $\omega$,
\begin{eqnarray}
   l_m \equiv (\mathbf{L}\mathbf{D}\mathbf{R})_{m0} = -\frac{4}{\pi^2} \sum_k  \frac{e^{j k \Omega \tau} e^{jk\theta}}{(m-k)k},
\end{eqnarray}
when $m$ is even.  By examination, we can then see that 
the spectral weight at the input frequency is
\begin{eqnarray}
    (\mathbf{L}\mathbf{D}\mathbf{R})_{00} = \frac{4}{\pi^2} \sum_k  \frac{e^{j k (\Omega \tau+\theta)}}{k^2},
\end{eqnarray}
which is $1$ when $\Omega \tau + \theta = 0$.  

The scattering matrix for the odd mode excitations $\mathbf{S_o}$ is therefore:
\begin{eqnarray}
    \mathbf{S_o} = \left(\begin{array}{cc} 
    0 & l_m \\  
    r_m & 0 \\ 
    \end{array}\right).
\end{eqnarray}
Note that this scattering matrix describes the way the network scatters an incident signal among its physical ports, as well as effective ports indexed by $m$, which represent signals at sidebands of the modulation rate $\Omega$.

Finally, to recover the full scattering matrix in the numbered port basis we perform the inverse transformation, $\mathbf{S} = \mathbf{U}^{-1} \mathbf{S}_{e,o} \mathbf{U}$.  Here, $\mathbf{S}_{e,o}$ is a block-diagonal matrix with diagonal entries $\mathbf{S}_e$ and $\mathbf{S}_o$.  This yields:
\begin{eqnarray}
    && \mathbf{S} = \frac{1}{2} \times \\
    && \left(\begin{array}{cccc} 
    0 & \delta_{m,0}+l_m & 0 & \delta_{m,0}-l_m \\  
    \delta_{m,0}-r_m & 0 & \delta_{m,0}+r_m & 0 \\  
    0 & \delta_{m,0}-l_m & 0 & \delta_{m,0}+l_m \\ 
    \delta_{m,0}+r_m & 0 & \delta_{m,0}-r_m & 0 \\ 
    \end{array}\right), \nonumber
    \label{Stheory}
\end{eqnarray}
which reduces to the ideal scattering matrix for a four-port circulator when the operation condition $\Omega \tau = \pi/2 = \pm \theta$ is satisfied.  It also provides additional information about how the network adds sidebands when this criterion is violated.

The above frequency-domain analysis makes a convenient starting-point for studies of group delay dispersion, frequency-dependent loss, and finite modulation bandwidth.  To account for group delay dispersion and frequency-depdendent loss, the delay matrix elements become
\begin{eqnarray}
    \mathbf{D}^{\prime}_{mn} = a_m e^{jm\Omega \tau_m} \delta_{m,n}.
    \label{delay}
\end{eqnarray}
The frequency-dependence of the delay and attenuation in the line are now accounted for by the terms $\tau_m$ and $a_m$.  

To handle finite modulation bandwidth, we assume that for ``intermediate'' biasing, the symmetric lattice is approximately balanced.  In this configuration, the even-mode scattering matrix is unchanged from Eq.~(\ref{Se}), and the odd-mode scattering matrix is perfectly reflective (the two-by-two identity matrix), with some overall undetermined phase.  The full network therefore scatters incident signals promptly, dividing their power equally between all ports.  In this limit, finite modulation bandwidth may be treated by truncating the sum in $k$.  

Together, these three effects give revised expressions for $l_m$ and $r_m$ which we denote $l^\prime_m$ and $r^\prime_m$:
\begin{eqnarray}
\label{pmprime} 
l^\prime_m = -\frac{4}{\pi^2} \sum_k^{k_{max}} a_k \frac{e^{j k \Omega \tau_k} e^{jk\theta}}{(m-k)k},  \\ \nonumber
r^\prime_m = -\frac{4}{\pi^2} \sum_k^{k_{max}} a_k \frac{e^{j k \Omega \tau_k} e^{j(m-k)\theta}}{(m-k)k}. 
\end{eqnarray}


When $k_\textrm{max}$ is finite, $l^\prime_m$ and $r^\prime_m$ have magnitudes which are less than $1$, and Eq.~(\ref{Stheory}) is amended in two ways.  First, the diagonal elements in the even and odd rows become $\tilde{l}_m/2$ and $\tilde{r}_m/2$, respectively.  These parameters are constrained by conservation of energy to have magnitudes $|\tilde{l}_m| = \sqrt{1-l_m^2}$ and $|\tilde{r}_m| = \sqrt{1-r_m^2}$. Second, the transmission elements which were perfectly vanishing become $-\tilde{l}_m/2$ in the odd rows and $-\tilde{r}_m/2$ in the even rows.

\begin{figure}[hbt]
\begin{center}
\includegraphics[width=1.0\columnwidth]{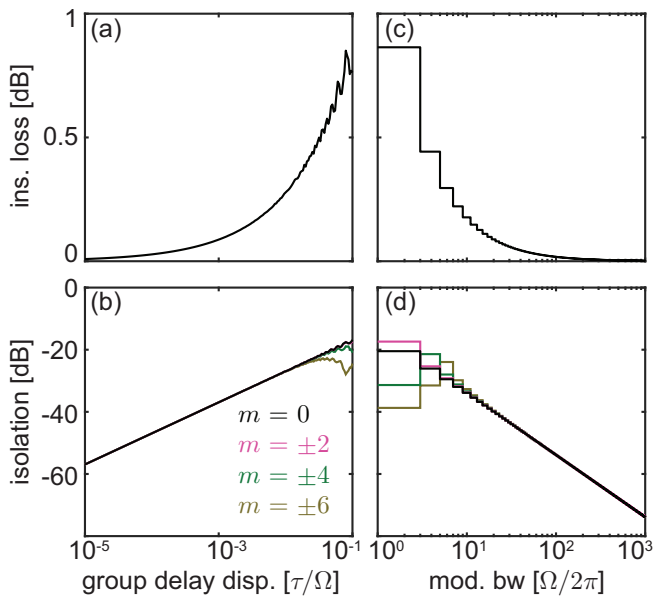}
\caption {\textbf{Performance in the presence of non-idealities.} The effects of group delay dispersion (a-b) and finite modulation bandwidth (c-d) on the circulator's insertion loss (a,c) and isolation (b,d), calculated with Eq.~(\ref{pmprime})
}
\label{fig:analyticnonidealities}
\end{center}
\end{figure}

Results from this revised expression are plotted in Fig.~\ref{fig:analyticnonidealities}, which shows the design's sensitivity to group delay dispersion in the network's delay lines (a-b) and the finite modulation bandwidth of its transfer switches (c-d).  For each case, the insertion loss (a,c) and isolation (b,d) are plotted.  

\section{Transfer switch scattering parameters}
\label{sec:switchS}
Here we derive an analytic expression for the scattering matrix of the four-port transfer switch in Fig.~\ref{fig:symlattice}a. This switch has two types of inductors, $l_t$ and $l_c$, along with capacitors $c$ and ports of characteristic impedance $Z_0$. To obtain the frequency dependent scattering matrix $\mathbf{S}[\omega]$, we first calculate the admittance matrix $\mathbf{Y}[\omega]$ of this circuit in the manner discussed in Ref.~\cite{nagappan:1970} and references therein: 
\begin{eqnarray}
    \mathbf{Y}[\omega] = \mathbf{A}^T \, \mathbf{y}[\omega] \, \mathbf{A},
    \label{Y_eqn}
\end{eqnarray}
where $\mathbf{A}$ is the incidence matrix describing nodal connectivity of the network, and $\mathbf{y}$ is the primitive admittance matrix describing the voltage current relationship across each chord of the network. 

The circuit described in Fig.~\ref{fig:symlattice}a has four nodes in addition to ground (each node labeled by the number of the port connected to it) and eight chords (chord numbers 1 through 4 across the inductors, and chord numbers 5 through 8 connecting the capacitors to ground). The incidence matrix $\mathbf{A}$ is therefore the $8 \times 4$ matrix,
\begin{equation}
\mathbf{A} = \begin{bmatrix}
    1 & -1 & 0 & 0 \\
    1 & 0 & 0 & -1 \\
    0 & -1 & 1 & 0 \\
    0 & 0 & 1 & -1 \\
    1 & 0 & 0 & 0 \\
    0 & 1 & 0 & 0 \\
    0 & 0 & 1 & 0 \\
    0 & 0 & 0 & 1 \\
    \end{bmatrix},
    \label{A_def}
\end{equation}
and the primitive admittance matrix is the $8 \times 8$ diagonal matrix,

\begin{equation}
\mathbf{y}[\omega] = \textrm{diag}\{
    \frac{1}{j \omega l_t}, \frac{1}{j \omega l_c}, \frac{1}{j \omega l_c}, \frac{1}{j \omega l_t}, j \omega c, j \omega c, j \omega c, j \omega c\}.
    \label{y_def}
\end{equation}
Substituting Eq.~(\ref{A_def}) and Eq.~(\ref{y_def}) into Eq.~(\ref{Y_eqn}) provides a straightforward construction of the admittance matrix, which is readily transformed into the four-port scattering matrix via the formula \cite{pozar:2011}:
\begin{equation}
    \mathbf{S}[\omega] = \left(\mathbf{I} + Z_0 \mathbf{Y}[\omega] \right)^{-1} \left(\mathbf{I} - Z_0 \mathbf{Y}[\omega] \right).
    \label{S_def}
\end{equation}
Here $\mathbf{I}$ is the $4 \times 4$ identity matrix. 

The analytic expression for $\mathbf{S}[\omega]$ can now be expanded in $\epsilon$, the small parameter which describes bridge imbalance.  We assume the switch is in its crossed position, and take $l_t = l_0/\epsilon$ and $l_c=l_0$. To further simplify the result, we also introduce the capacitive time constant $\tau_c = Z_0 c$. The network described in Fig.~\ref{fig:symlattice}a then functions as a matched crossover switch when the condition $l_0/(2c) = Z_0^2 / \left(1 + \omega^2 \tau_c^2 \right)$ is met---that is, when the impedance of the resonator $\sqrt{l_0/(2c)}$ equals the characteristic impedance of the surrounding lines $Z_0$, and $\omega^2 \tau_c^2 \ll 1$. Enforcing this condition, the first column of the switch's scattering matrix is,
\begin{eqnarray}
   |\mathbf{S}_{11}| &=& \epsilon \left( \frac{|\tau_c^2 \omega^2 - 1|}{2 \tau_c \omega} \right) + \mathcal{O}(\epsilon^2), \nonumber \\
   |\mathbf{S}_{21}| &=& |\mathbf{S}_{11}|, \nonumber \\
   |\mathbf{S}_{31}| &=& \epsilon \left( \frac{|\tau_c \omega + j|^2}{2 \tau_c \omega} \right) + \mathcal{O}(\epsilon^2), \nonumber \\
   |\mathbf{S}_{41}| &=& \left\vert \frac{j + \tau_c \omega}{j - \tau_c \omega} - j \epsilon \left( \frac{(\tau_c \omega + j)^2}{2 \tau_c \omega} \right) \right\vert + \mathcal{O}(\epsilon^2). \nonumber \\
   \label{epsexpansion}
\end{eqnarray}
Eq.~(\ref{epsexpansion}) shows that the switch's isolation and return loss scale linearly with $\epsilon$ in the limit of large imbalance.

\section{Widely tunable dc-SQUID inductors}
\label{sec:SQUIDs}

\begin{figure}
\begin{center}
\includegraphics[width=1.0\columnwidth]{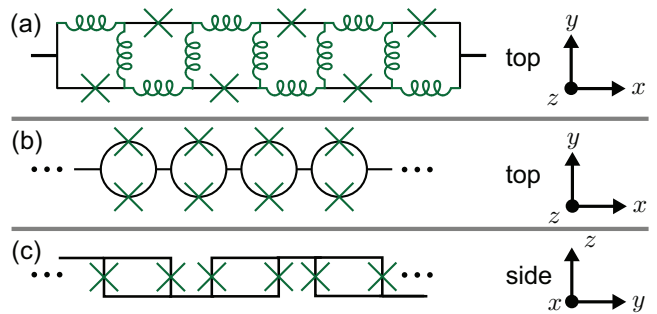}
\caption {\textbf{Tunable inductor design.} 
(a) Top view of a ladder of alternating rf-SQUIDs, as studied in Refs.~\cite{bell:2012,naaman:2017}. The linear inductors may be realized either with the geometric inductance of galvanic connections, or with high critical current Josephson junctions. When biased close to frustration the impedance of the array can exceed a resistance quantum~\cite{bell:2012}.  
(b) Top view of a series-array of dc-SQUIDs, with SQUID loops in the plane of the chip. The tunability of such arrays may be limited by the geometric inductance of the SQUID loops. 
(c) Side view of a series array of dc-SQUIDs, with SQUID loops orthogonal to the plane of the chip.  In this "vertical" geometry, the vertical dimension of the loop is set by the inter-layer spacing, which can substantially reduce the loop's geometric inductance~\cite{hover:2012}.}
\label{fig:SQUIDstyles}
\end{center}
\end{figure}

In this section we discuss two ways to make the highly tunable inductors needed for the transfer switch.  The first uses laddered arrays of Josephson junctions, as depicted in Fig.~\ref{fig:SQUIDstyles}a.  Inductors of this type are demonstrated and discussed in Refs.~\cite{bell:2012,naaman:2017}.

The second uses arrays of dc-SQUIDs formed by two Josephson junctions arranged in parallel. The tunability of a dc-SQUID may be limited by a) the degree to which the critical currents of the two junctions may be made identical, and b) the geometric inductance of the loop~\cite{vanduzer:1981}. Cascading an array of these elements in series dilutes the nonlinearity by decreasing the superconducting phase drop across each junction, thus increasing power-handling~\cite{kerckhoff:2015}.

In an optical lithographic process where the SQUID is fabricated in the plane of the chip (see schematic in Fig.~\ref{fig:SQUIDstyles}b), such a loop can be made no smaller than several microns on a side.  For Josephson junctions with critical currents on the scale of 10 $\mu$A, the geometric inductance of such a loop is often the limitation on tunability.

That limitation can be removed by orienting the SQUID ``vertically'', such that a vector normal to its loop is parallel with the chip's surface (Fig.~\ref{fig:SQUIDstyles}c). The small dimension of the loop is now determined by the inter-layer separation ($\approx 100$ nm), rather than the minimum feature size in the optical lithographic process.  This decreases the volume in which the loop can store magnetic energy by a factor of $\approx50$, reducing its geometric inductance by the same amount~\cite{hover:2012}. 

Critical current symmetry is a process dependent quantity, but for niobium trilayer processes variations are typically below 5\%~\cite{tolpygo:2017}. Subject to that limitation, a vertical SQUID with nominal junction critical currents of $10$~$\mu$A  could be tuned by a factor of 40.  This is the basis for our estimate that $\epsilon \leq 2.5 \times 10^{-2}$ would be possible with such a design.

\section{Cross-talk and added noise}
\label{sec:noise}
In principle, a circulator based on delays and dynamically modulated switches need not add noise.  Firstly, an ideal circulator functions with unit gain, so noiseless circulation is permitted by the fundamental theorem for phase-insensitive amplifiers~\cite{caves:1982}.  
Secondly, dynamically modulated switches may route signals without adding noise~\cite{yurke:1984}.  Nevertheless, the proposed circulator is an active, flux-actuated device.  It may therefore add noise via the fluctuations of its control flux.  

These variations around the optimal flux bias point may cause stochastic changes in the circulator's scattering parameters.  The result is a sort of multiplicative noise on the circulated signal.  This may be mitigated by thermalizing the bias lines to cryogenic temperatures, where the Johnson noise is small relative to the $\approx 100$ $\mu$A bias currents that generate the flux controls.  In similar flux-modulated circulators, thermalization at $4$ K keeps added noise below half a photon; further reduction is possible with filtering or thermalization to lower temperatures~\cite{chapman:2017b}.



\begin{figure}
\begin{center}
\includegraphics[width=1.0\columnwidth]{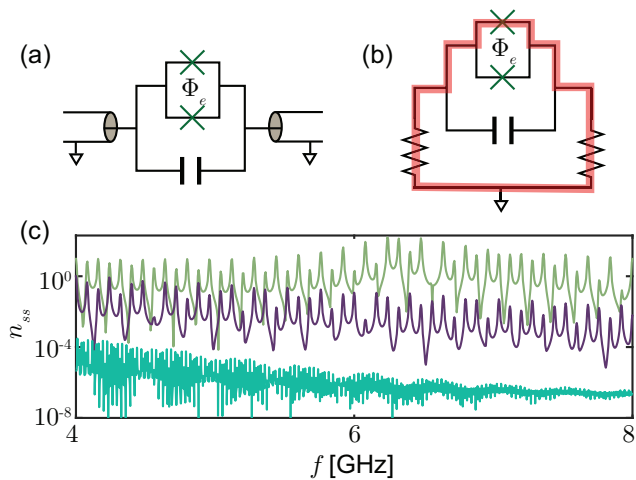}
\caption {\textbf{Noise from active flux control.} (a) Circuit model for noise from a dynamically modulated flux $\Phi_e$. (b) Alternate representation for the circuit in (a).  The pink path traces an Amperian loop which encloses a time-varying flux.  The resulting electromotive force drives current through the resistors.  (c) Steady-state photon number vs photon frequency for the circuit in (b), calculated with Eq.~(\ref{nss}).  The three traces show the resulting occupations when the external flux $\Phi_e(t)$ is modulated with a Fourier-series approximation of an $\Omega = 2\pi \times 80$ MHz square wave capped at $2$ GHz (olive), a sigma approximation~\cite{lanczos:1988} of an $\Omega = 2 \pi \times 80$ MHz square wave capped at $2$ GHz (purple), and a sigma approximation of an $\Omega = 2 \pi \times 15$ MHz square wave capped at $750$ MHz (teal).}
\label{fig:noise}
\end{center}
\end{figure}

Cross-talk between the bias lines and the microwave circuit can also create unwanted photons at the circulator's output.  Effects caused by spurious coupling may be reduced with careful layout, such as designing flux lines that are higher-order multipole sources of magnetic field, to decrease the range over which stray fluxes are broadcast.  But even by itself, the intentional coupling engineered between the bias lines and the loops of the SQUIDs may be problematic, as it can allow the time-varying control flux to drive current out of the circulator's ports. Fig.~\ref{fig:noise}a illustrates how this can happen in a simple model comprised of a dc SQUID with self-capacitance $c_J$, inserted into a transmission line of characteristic impedance $Z_0$.  As each transmission line presents a $Z_0$ path to ground, there exist finite-impedance Amperian loops that carry current out of the circuit when $d\Phi_e/dt \neq 0$ (see Fig.~\ref{fig:noise}b). 

In this manner, the flux bias can create photons at a range of frequencies.  To quantify this effect, we numerically integrate the classical equation of motion for the circuit in Fig.~\ref{fig:noise}b~\cite{vanduzer:1981}, and Fourier transform its steady-state dynamics to obtain the power spectral density $p(f)$.  We fix the junction assymmetry at 5\%, neglect the geometric inductance of the junctions, and take the the capacitance to be $c_J = 180$ fF and the quality factor to be $Q = 3.5$. Power-balance then allows expression of the steady-state photon occupation $n_{ss}$ in a cavity irradiated by this noise source:
\begin{eqnarray}
   n_{ss} \leq \frac{p(f)}{\hbar \omega}.
   \label{nss}
\end{eqnarray}
We write Eq.~(\ref{nss}) as an inequality to indicate that it is a kind of worst-case scenario: if any of the flux $\Phi_e$ is returned within the Amperian loop drawn in Fig.~\ref{fig:noise}b, the irradiated noise will decrease; it vanishes completely in the limit where the flux is returned symmetrically within (and outside of) the Amperian loop.

Fig.~\ref{fig:noise}c shows the steady-state photon number at frequencies between 4 and 8 GHz, for a flux signal comprised of the first 25 terms in the Fourier series of an $\Omega = 2 \pi \times 80$ MHz square wave (blue trace).  The presence of non-zero occupation at frequencies above the largest spectral component of the modulation ($2$ GHz) is a signature of the SQUID's nonlinearity.

Photon population is well above the $\approx10^{-4}$ threshold for residual cavity occupation now being approached in circuit QED experiments~\cite{yeh:2017,yan:2018,wang:2018}, but this may be improved by regulation of the flux control signal's spectral content.  The purple trace in Fig.~\ref{fig:noise}d shows the same measurement, now performed with a sigma approximation~\cite{lanczos:1988} of the square wave---essentially a low-pass filtered square wave.  The resulting occupancies are suppressed by several orders of magnitude.  Reducing the modulation rate to $\Omega = 2 \pi \times 15$ MHz and capping the modulation bandwidth at $750$ MHz provides an additional reduction in the occupancy (teal trace).  The price of this filtering is paid in loss.  
Because the switching time is increased by the $750$ MHz cap on the modulation bandwidth, insertion loss climbs to $-0.5$ dB.  

Ultimately, the above analysis is meant to model how a junction's nonlinearity can mix flux control signals up in frequency, and how these microwave photons are emitted out of the circulator's ports.  This radiation can be controlled by regulating the modulation spectrum, and with symmetric arrangement of the bias lines in layout.  Nevertheless, this model is merely an approximation: we discuss it here to illustrate possible recourse.


\begin{thebibliography}{46}%
\makeatletter
\providecommand \@ifxundefined [1]{%
 \@ifx{#1\undefined}
}%
\providecommand \@ifnum [1]{%
 \ifnum #1\expandafter \@firstoftwo
 \else \expandafter \@secondoftwo
 \fi
}%
\providecommand \@ifx [1]{%
 \ifx #1\expandafter \@firstoftwo
 \else \expandafter \@secondoftwo
 \fi
}%
\providecommand \natexlab [1]{#1}%
\providecommand \enquote  [1]{``#1''}%
\providecommand \bibnamefont  [1]{#1}%
\providecommand \bibfnamefont [1]{#1}%
\providecommand \citenamefont [1]{#1}%
\providecommand \href@noop [0]{\@secondoftwo}%
\providecommand \href [0]{\begingroup \@sanitize@url \@href}%
\providecommand \@href[1]{\@@startlink{#1}\@@href}%
\providecommand \@@href[1]{\endgroup#1\@@endlink}%
\providecommand \@sanitize@url [0]{\catcode `\\12\catcode `\$12\catcode
  `\&12\catcode `\#12\catcode `\^12\catcode `\_12\catcode `\%12\relax}%
\providecommand \@@startlink[1]{}%
\providecommand \@@endlink[0]{}%
\providecommand \url  [0]{\begingroup\@sanitize@url \@url }%
\providecommand \@url [1]{\endgroup\@href {#1}{\urlprefix }}%
\providecommand \urlprefix  [0]{URL }%
\providecommand \Eprint [0]{\href }%
\providecommand \doibase [0]{http://dx.doi.org/}%
\providecommand \selectlanguage [0]{\@gobble}%
\providecommand \bibinfo  [0]{\@secondoftwo}%
\providecommand \bibfield  [0]{\@secondoftwo}%
\providecommand \translation [1]{[#1]}%
\providecommand \BibitemOpen [0]{}%
\providecommand \bibitemStop [0]{}%
\providecommand \bibitemNoStop [0]{.\EOS\space}%
\providecommand \EOS [0]{\spacefactor3000\relax}%
\providecommand \BibitemShut  [1]{\csname bibitem#1\endcsname}%
\let\auto@bib@innerbib\@empty
\bibitem [{\citenamefont {Sliwa}\ \emph {et~al.}(2015)\citenamefont {Sliwa},
  \citenamefont {Hatridge}, \citenamefont {Narla}, \citenamefont {Shankar},
  \citenamefont {Frunzio}, \citenamefont {Schoelkopf},\ and\ \citenamefont
  {Devoret}}]{sliwa:2015}%
  \BibitemOpen
  \bibfield  {author} {\bibinfo {author} {\bibfnamefont {K.~M.}\ \bibnamefont
  {Sliwa}}, \bibinfo {author} {\bibfnamefont {M}~\bibnamefont {Hatridge}},
  \bibinfo {author} {\bibfnamefont {A}~\bibnamefont {Narla}}, \bibinfo {author}
  {\bibfnamefont {S}~\bibnamefont {Shankar}}, \bibinfo {author} {\bibfnamefont
  {L}~\bibnamefont {Frunzio}}, \bibinfo {author} {\bibfnamefont {R.~J.}\
  \bibnamefont {Schoelkopf}}, \ and\ \bibinfo {author} {\bibfnamefont {M.~H.}\
  \bibnamefont {Devoret}},\ }\bibfield  {title} {\enquote {\bibinfo {title}
  {Reconfigurable {J}osephson circulator/directional amplifier},}\ }\href@noop
  {} {\bibfield  {journal} {\bibinfo  {journal} {Phys. Rev. X}\ }\textbf
  {\bibinfo {volume} {5}},\ \bibinfo {pages} {041020} (\bibinfo {year}
  {2015})}\BibitemShut {NoStop}%
\bibitem [{\citenamefont {Lecocq}\ \emph {et~al.}(2017)\citenamefont {Lecocq},
  \citenamefont {Ranzani}, \citenamefont {Peterson}, \citenamefont {Cicak},
  \citenamefont {Simmonds}, \citenamefont {Teufel},\ and\ \citenamefont
  {Aumentado}}]{lecocq:2017}%
  \BibitemOpen
  \bibfield  {author} {\bibinfo {author} {\bibfnamefont {F.}~\bibnamefont
  {Lecocq}}, \bibinfo {author} {\bibfnamefont {L.}~\bibnamefont {Ranzani}},
  \bibinfo {author} {\bibfnamefont {G.~A.}\ \bibnamefont {Peterson}}, \bibinfo
  {author} {\bibfnamefont {K.}~\bibnamefont {Cicak}}, \bibinfo {author}
  {\bibfnamefont {R.~W.}\ \bibnamefont {Simmonds}}, \bibinfo {author}
  {\bibfnamefont {J.~D.}\ \bibnamefont {Teufel}}, \ and\ \bibinfo {author}
  {\bibfnamefont {J.}~\bibnamefont {Aumentado}},\ }\bibfield  {title} {\enquote
  {\bibinfo {title} {Nonreciprocal microwave signal processing with a
  field-programmable {J}osephson amplifier},}\ }\href {\doibase
  10.1103/PhysRevApplied.7.024028} {\bibfield  {journal} {\bibinfo  {journal}
  {Phys. Rev. Applied}\ }\textbf {\bibinfo {volume} {7}},\ \bibinfo {pages}
  {024028} (\bibinfo {year} {2017})}\BibitemShut {NoStop}%
\bibitem [{\citenamefont {Chapman}\ \emph
  {et~al.}(2017{\natexlab{a}})\citenamefont {Chapman}, \citenamefont
  {Rosenthal}, \citenamefont {Kerckhoff}, \citenamefont {Moores}, \citenamefont
  {Vale}, \citenamefont {Mates}, \citenamefont {Hilton}, \citenamefont
  {Lalumi\`ere}, \citenamefont {Blais},\ and\ \citenamefont
  {Lehnert}}]{chapman:2017b}%
  \BibitemOpen
  \bibfield  {author} {\bibinfo {author} {\bibfnamefont {Benjamin~J.}\
  \bibnamefont {Chapman}}, \bibinfo {author} {\bibfnamefont {Eric~I.}\
  \bibnamefont {Rosenthal}}, \bibinfo {author} {\bibfnamefont {Joseph}\
  \bibnamefont {Kerckhoff}}, \bibinfo {author} {\bibfnamefont {Bradley~A.}\
  \bibnamefont {Moores}}, \bibinfo {author} {\bibfnamefont {Leila~R.}\
  \bibnamefont {Vale}}, \bibinfo {author} {\bibfnamefont {J.~A.~B.}\
  \bibnamefont {Mates}}, \bibinfo {author} {\bibfnamefont {Gene~C.}\
  \bibnamefont {Hilton}}, \bibinfo {author} {\bibfnamefont {Kevin}\
  \bibnamefont {Lalumi\`ere}}, \bibinfo {author} {\bibfnamefont {Alexandre}\
  \bibnamefont {Blais}}, \ and\ \bibinfo {author} {\bibfnamefont {K.~W.}\
  \bibnamefont {Lehnert}},\ }\bibfield  {title} {\enquote {\bibinfo {title}
  {Widely tunable on-chip microwave circulator for superconducting quantum
  circuits},}\ }\href {\doibase 10.1103/PhysRevX.7.041043} {\bibfield
  {journal} {\bibinfo  {journal} {Phys. Rev. X}\ }\textbf {\bibinfo {volume}
  {7}},\ \bibinfo {pages} {041043} (\bibinfo {year}
  {2017}{\natexlab{a}})}\BibitemShut {NoStop}%
\bibitem [{\citenamefont {Macklin}\ \emph {et~al.}(2015)\citenamefont
  {Macklin}, \citenamefont {O`Brien}, \citenamefont {Hover}, \citenamefont
  {Schwartz}, \citenamefont {Bolkhovsky}, \citenamefont {Zhang}, \citenamefont
  {Oliver},\ and\ \citenamefont {Siddiqi}}]{macklin:2015}%
  \BibitemOpen
  \bibfield  {author} {\bibinfo {author} {\bibfnamefont {C}~\bibnamefont
  {Macklin}}, \bibinfo {author} {\bibfnamefont {K}~\bibnamefont {O`Brien}},
  \bibinfo {author} {\bibfnamefont {D}~\bibnamefont {Hover}}, \bibinfo {author}
  {\bibfnamefont {M.~E.}\ \bibnamefont {Schwartz}}, \bibinfo {author}
  {\bibfnamefont {V}~\bibnamefont {Bolkhovsky}}, \bibinfo {author}
  {\bibfnamefont {X}~\bibnamefont {Zhang}}, \bibinfo {author} {\bibfnamefont
  {W.~D.}\ \bibnamefont {Oliver}}, \ and\ \bibinfo {author} {\bibfnamefont
  {I}~\bibnamefont {Siddiqi}},\ }\bibfield  {title} {\enquote {\bibinfo {title}
  {A near--quantum-limited {J}osephson traveling-wave parametric amplifier},}\
  }\href@noop {} {\bibfield  {journal} {\bibinfo  {journal} {Science}\ }\textbf
  {\bibinfo {volume} {350}},\ \bibinfo {pages} {307--310} (\bibinfo {year}
  {2015})}\BibitemShut {NoStop}%
\bibitem [{\citenamefont {Roy}\ \emph {et~al.}(2015)\citenamefont {Roy},
  \citenamefont {Kundu}, \citenamefont {Chand}, \citenamefont {Vadiraj},
  \citenamefont {Ranadive}, \citenamefont {Nehra}, \citenamefont {Patankar},
  \citenamefont {Aumentado}, \citenamefont {Clerk},\ and\ \citenamefont
  {Vijay}}]{roy:2015}%
  \BibitemOpen
  \bibfield  {author} {\bibinfo {author} {\bibfnamefont {Tanay}\ \bibnamefont
  {Roy}}, \bibinfo {author} {\bibfnamefont {Suman}\ \bibnamefont {Kundu}},
  \bibinfo {author} {\bibfnamefont {Madhavi}\ \bibnamefont {Chand}}, \bibinfo
  {author} {\bibfnamefont {A.~M.}\ \bibnamefont {Vadiraj}}, \bibinfo {author}
  {\bibfnamefont {A.}~\bibnamefont {Ranadive}}, \bibinfo {author}
  {\bibfnamefont {N.}~\bibnamefont {Nehra}}, \bibinfo {author} {\bibfnamefont
  {Meghan~P.}\ \bibnamefont {Patankar}}, \bibinfo {author} {\bibfnamefont
  {J.}~\bibnamefont {Aumentado}}, \bibinfo {author} {\bibfnamefont {A.~A.}\
  \bibnamefont {Clerk}}, \ and\ \bibinfo {author} {\bibfnamefont
  {R.}~\bibnamefont {Vijay}},\ }\bibfield  {title} {\enquote {\bibinfo {title}
  {Broadband parametric amplification with impedance engineering: Beyond the
  gain-bandwidth product},}\ }\href {\doibase 10.1063/1.4939148} {\bibfield
  {journal} {\bibinfo  {journal} {Applied Physics Letters}\ }\textbf {\bibinfo
  {volume} {107}},\ \bibinfo {pages} {262601} (\bibinfo {year} {2015})},\
  \Eprint {http://arxiv.org/abs/https://doi.org/10.1063/1.4939148}
  {https://doi.org/10.1063/1.4939148} \BibitemShut {NoStop}%
\bibitem [{\citenamefont {Naaman}\ \emph
  {et~al.}(2017{\natexlab{a}})\citenamefont {Naaman}, \citenamefont
  {Ferguson},\ and\ \citenamefont {Epstein}}]{naaman:2017}%
  \BibitemOpen
  \bibfield  {author} {\bibinfo {author} {\bibfnamefont {O}~\bibnamefont
  {Naaman}}, \bibinfo {author} {\bibfnamefont {D.~G.}\ \bibnamefont
  {Ferguson}}, \ and\ \bibinfo {author} {\bibfnamefont {R.~J.}\ \bibnamefont
  {Epstein}},\ }\bibfield  {title} {\enquote {\bibinfo {title} {High saturation
  power {J}osephson parametric amplifier with {G}{H}z bandwidth},}\ }\href@noop
  {} {\bibfield  {journal} {\bibinfo  {journal} {arXiv preprint
  arXiv:1711.07549}\ } (\bibinfo {year} {2017}{\natexlab{a}})}\BibitemShut
  {NoStop}%
\bibitem [{\citenamefont {Abdo}\ \emph {et~al.}(2017)\citenamefont {Abdo},
  \citenamefont {Brink},\ and\ \citenamefont {Chow}}]{abdo:2017}%
  \BibitemOpen
  \bibfield  {author} {\bibinfo {author} {\bibfnamefont {Baleegh}\ \bibnamefont
  {Abdo}}, \bibinfo {author} {\bibfnamefont {Markus}\ \bibnamefont {Brink}}, \
  and\ \bibinfo {author} {\bibfnamefont {Jerry~M.}\ \bibnamefont {Chow}},\
  }\bibfield  {title} {\enquote {\bibinfo {title} {Gyrator operation using
  {J}osephson mixers},}\ }\href {\doibase 10.1103/PhysRevApplied.8.034009}
  {\bibfield  {journal} {\bibinfo  {journal} {Phys. Rev. Applied}\ }\textbf
  {\bibinfo {volume} {8}},\ \bibinfo {pages} {034009} (\bibinfo {year}
  {2017})}\BibitemShut {NoStop}%
\bibitem [{\citenamefont {Fay}\ and\ \citenamefont
  {Comstock}(1965)}]{fay:1965}%
  \BibitemOpen
  \bibfield  {author} {\bibinfo {author} {\bibfnamefont {C.~E.}\ \bibnamefont
  {Fay}}\ and\ \bibinfo {author} {\bibfnamefont {R.~L.}\ \bibnamefont
  {Comstock}},\ }\bibfield  {title} {\enquote {\bibinfo {title} {Operation of
  the ferrite junction circulator},}\ }\href {\doibase
  10.1109/TMTT.1965.1125923} {\bibfield  {journal} {\bibinfo  {journal} {IEEE
  Transactions on Microwave Theory and Techniques}\ }\textbf {\bibinfo {volume}
  {13}},\ \bibinfo {pages} {15--27} (\bibinfo {year} {1965})}\BibitemShut
  {NoStop}%
\bibitem [{\citenamefont {Fowler}\ \emph {et~al.}(2012)\citenamefont {Fowler},
  \citenamefont {Mariantoni}, \citenamefont {Martinis},\ and\ \citenamefont
  {Cleland}}]{fowler:2012}%
  \BibitemOpen
  \bibfield  {author} {\bibinfo {author} {\bibfnamefont {Austin~G.}\
  \bibnamefont {Fowler}}, \bibinfo {author} {\bibfnamefont {Matteo}\
  \bibnamefont {Mariantoni}}, \bibinfo {author} {\bibfnamefont {John~M.}\
  \bibnamefont {Martinis}}, \ and\ \bibinfo {author} {\bibfnamefont
  {Andrew~N.}\ \bibnamefont {Cleland}},\ }\bibfield  {title} {\enquote
  {\bibinfo {title} {Surface codes: Towards practical large-scale quantum
  computation},}\ }\href {\doibase 10.1103/PhysRevA.86.032324} {\bibfield
  {journal} {\bibinfo  {journal} {Phys. Rev. A}\ }\textbf {\bibinfo {volume}
  {86}},\ \bibinfo {pages} {032324} (\bibinfo {year} {2012})}\BibitemShut
  {NoStop}%
\bibitem [{\citenamefont {Galland}\ \emph {et~al.}(2013)\citenamefont
  {Galland}, \citenamefont {Ding}, \citenamefont {Harris}, \citenamefont
  {Baehr-Jones},\ and\ \citenamefont {Hochberg}}]{galland:2013}%
  \BibitemOpen
  \bibfield  {author} {\bibinfo {author} {\bibfnamefont {Christophe}\
  \bibnamefont {Galland}}, \bibinfo {author} {\bibfnamefont {Ran}\ \bibnamefont
  {Ding}}, \bibinfo {author} {\bibfnamefont {Nicholas~C}\ \bibnamefont
  {Harris}}, \bibinfo {author} {\bibfnamefont {Tom}\ \bibnamefont
  {Baehr-Jones}}, \ and\ \bibinfo {author} {\bibfnamefont {Michael}\
  \bibnamefont {Hochberg}},\ }\bibfield  {title} {\enquote {\bibinfo {title}
  {Broadband on-chip optical non-reciprocity using phase modulators},}\
  }\href@noop {} {\bibfield  {journal} {\bibinfo  {journal} {Optics express}\
  }\textbf {\bibinfo {volume} {21}},\ \bibinfo {pages} {14500--14511} (\bibinfo
  {year} {2013})}\BibitemShut {NoStop}%
\bibitem [{\citenamefont {Yang}\ \emph {et~al.}(2014)\citenamefont {Yang},
  \citenamefont {Galland}, \citenamefont {Liu}, \citenamefont {Tan},
  \citenamefont {Ding}, \citenamefont {Li}, \citenamefont {Bergman},
  \citenamefont {Baehr-Jones},\ and\ \citenamefont {Hochberg}}]{yang:2014}%
  \BibitemOpen
  \bibfield  {author} {\bibinfo {author} {\bibfnamefont {Yisu}\ \bibnamefont
  {Yang}}, \bibinfo {author} {\bibfnamefont {Christophe}\ \bibnamefont
  {Galland}}, \bibinfo {author} {\bibfnamefont {Yang}\ \bibnamefont {Liu}},
  \bibinfo {author} {\bibfnamefont {Kang}\ \bibnamefont {Tan}}, \bibinfo
  {author} {\bibfnamefont {Ran}\ \bibnamefont {Ding}}, \bibinfo {author}
  {\bibfnamefont {Qi}~\bibnamefont {Li}}, \bibinfo {author} {\bibfnamefont
  {Keren}\ \bibnamefont {Bergman}}, \bibinfo {author} {\bibfnamefont {Tom}\
  \bibnamefont {Baehr-Jones}}, \ and\ \bibinfo {author} {\bibfnamefont
  {Michael}\ \bibnamefont {Hochberg}},\ }\bibfield  {title} {\enquote {\bibinfo
  {title} {Experimental demonstration of broadband {L}orentz non-reciprocity in
  an integrable photonic architecture based on {M}ach-{Z}ehnder modulators},}\
  }\href@noop {} {\bibfield  {journal} {\bibinfo  {journal} {Optics express}\
  }\textbf {\bibinfo {volume} {22}},\ \bibinfo {pages} {17409--17422} (\bibinfo
  {year} {2014})}\BibitemShut {NoStop}%
\bibitem [{\citenamefont {Chapman}(2017)}]{chapman:2017c}%
  \BibitemOpen
  \bibfield  {author} {\bibinfo {author} {\bibfnamefont {Benjamin~J.}\
  \bibnamefont {Chapman}},\ }\emph {\bibinfo {title} {Widely tunable on-chip
  microwave circulator for superconducting quantum circuits}},\ \href@noop {}
  {Ph.D. thesis},\ \bibinfo  {school} {University of Colorado, Boulder}
  (\bibinfo {year} {2017}),\ \bibinfo {note} {chapter 7.}\BibitemShut {Stop}%
\bibitem [{\citenamefont {Lu}\ \emph {et~al.}(2018)\citenamefont {Lu},
  \citenamefont {Krol}, \citenamefont {Gao},\ and\ \citenamefont
  {Gong}}]{lu:2018a}%
  \BibitemOpen
  \bibfield  {author} {\bibinfo {author} {\bibfnamefont {Ruochen}\ \bibnamefont
  {Lu}}, \bibinfo {author} {\bibfnamefont {John}\ \bibnamefont {Krol}},
  \bibinfo {author} {\bibfnamefont {Liuqing}\ \bibnamefont {Gao}}, \ and\
  \bibinfo {author} {\bibfnamefont {Songbin}\ \bibnamefont {Gong}},\ }\bibfield
   {title} {\enquote {\bibinfo {title} {Frequency independent framework for
  synthesis of programmable non-reciprocal networks},}\ }\href@noop {}
  {\bibfield  {journal} {\bibinfo  {journal} {arXiv preprint arXiv:1801.01548}\
  } (\bibinfo {year} {2018})}\BibitemShut {NoStop}%
\bibitem [{\citenamefont {Biedka}\ \emph {et~al.}(2018)\citenamefont {Biedka},
  \citenamefont {Wu}, \citenamefont {Zou}, \citenamefont {Qin},\ and\
  \citenamefont {Wang}}]{biedka:2018}%
  \BibitemOpen
  \bibfield  {author} {\bibinfo {author} {\bibfnamefont {Mathew}\ \bibnamefont
  {Biedka}}, \bibinfo {author} {\bibfnamefont {Qianteng}\ \bibnamefont {Wu}},
  \bibinfo {author} {\bibfnamefont {Xiating}\ \bibnamefont {Zou}}, \bibinfo
  {author} {\bibfnamefont {Shihan}\ \bibnamefont {Qin}}, \ and\ \bibinfo
  {author} {\bibfnamefont {Yuanxun~Ethan}\ \bibnamefont {Wang}},\ }\bibfield
  {title} {\enquote {\bibinfo {title} {Integrated time-varying electromagnetic
  devices for ultra-wide band nonreciprocity},}\ }in\ \href@noop {} {\emph
  {\bibinfo {booktitle} {Radio and Wireless Symposium (RWS), 2018 IEEE}}}\
  (\bibinfo {organization} {IEEE},\ \bibinfo {year} {2018})\ pp.\ \bibinfo
  {pages} {80--83}\BibitemShut {NoStop}%
\bibitem [{\citenamefont {Reiskarimian}\ and\ \citenamefont
  {Krishnaswamy}(2016)}]{reiskarimian:2016}%
  \BibitemOpen
  \bibfield  {author} {\bibinfo {author} {\bibfnamefont {Negar}\ \bibnamefont
  {Reiskarimian}}\ and\ \bibinfo {author} {\bibfnamefont {Harish}\ \bibnamefont
  {Krishnaswamy}},\ }\bibfield  {title} {\enquote {\bibinfo {title}
  {Magnetic-free non-reciprocity based on staggered commutation},}\ }\href@noop
  {} {\bibfield  {journal} {\bibinfo  {journal} {Nature Communications}\
  }\textbf {\bibinfo {volume} {7}} (\bibinfo {year} {2016})}\BibitemShut
  {NoStop}%
\bibitem [{\citenamefont {Biedka}\ \emph {et~al.}(2017)\citenamefont {Biedka},
  \citenamefont {Zhu}, \citenamefont {Xu},\ and\ \citenamefont
  {Wang}}]{biedka:2017}%
  \BibitemOpen
  \bibfield  {author} {\bibinfo {author} {\bibfnamefont {Mathew~M}\
  \bibnamefont {Biedka}}, \bibinfo {author} {\bibfnamefont {Rui}\ \bibnamefont
  {Zhu}}, \bibinfo {author} {\bibfnamefont {Qiang~Mark}\ \bibnamefont {Xu}}, \
  and\ \bibinfo {author} {\bibfnamefont {Yuanxun~Ethan}\ \bibnamefont {Wang}},\
  }\bibfield  {title} {\enquote {\bibinfo {title} {Ultra-wide band
  non-reciprocity through sequentially-switched delay lines},}\ }\href@noop {}
  {\bibfield  {journal} {\bibinfo  {journal} {Scientific reports}\ }\textbf
  {\bibinfo {volume} {7}},\ \bibinfo {pages} {40014} (\bibinfo {year}
  {2017})}\BibitemShut {NoStop}%
\bibitem [{\citenamefont {Dinc}\ \emph {et~al.}(2017)\citenamefont {Dinc},
  \citenamefont {Tymchenko}, \citenamefont {Nagulu}, \citenamefont {Sounas},
  \citenamefont {Al\`u},\ and\ \citenamefont {Krishnaswamy}}]{dinc:2017}%
  \BibitemOpen
  \bibfield  {author} {\bibinfo {author} {\bibfnamefont {Tolga}\ \bibnamefont
  {Dinc}}, \bibinfo {author} {\bibfnamefont {Mykhailo}\ \bibnamefont
  {Tymchenko}}, \bibinfo {author} {\bibfnamefont {Aravind}\ \bibnamefont
  {Nagulu}}, \bibinfo {author} {\bibfnamefont {Dimitrios}\ \bibnamefont
  {Sounas}}, \bibinfo {author} {\bibfnamefont {Andrea}\ \bibnamefont {Al\`u}},
  \ and\ \bibinfo {author} {\bibfnamefont {Harish}\ \bibnamefont
  {Krishnaswamy}},\ }\bibfield  {title} {\enquote {\bibinfo {title}
  {Synchronized conductivity modulation to realize broadband lossless
  magnetic-free non-reciprocity},}\ }\href@noop {} {\bibfield  {journal}
  {\bibinfo  {journal} {Nature communications}\ }\textbf {\bibinfo {volume}
  {8}},\ \bibinfo {pages} {795} (\bibinfo {year} {2017})}\BibitemShut {NoStop}%
\bibitem [{\citenamefont {Rosenthal}\ \emph {et~al.}(2017)\citenamefont
  {Rosenthal}, \citenamefont {Chapman}, \citenamefont {Higginbotham},
  \citenamefont {Kerckhoff},\ and\ \citenamefont {Lehnert}}]{rosenthal:2017}%
  \BibitemOpen
  \bibfield  {author} {\bibinfo {author} {\bibfnamefont {Eric~I.}\ \bibnamefont
  {Rosenthal}}, \bibinfo {author} {\bibfnamefont {Benjamin~J.}\ \bibnamefont
  {Chapman}}, \bibinfo {author} {\bibfnamefont {Andrew~P.}\ \bibnamefont
  {Higginbotham}}, \bibinfo {author} {\bibfnamefont {Joseph}\ \bibnamefont
  {Kerckhoff}}, \ and\ \bibinfo {author} {\bibfnamefont {K.~W.}\ \bibnamefont
  {Lehnert}},\ }\bibfield  {title} {\enquote {\bibinfo {title} {Breaking
  {L}orentz reciprocity with frequency conversion and delay},}\ }\href
  {\doibase 10.1103/PhysRevLett.119.147703} {\bibfield  {journal} {\bibinfo
  {journal} {Phys. Rev. Lett.}\ }\textbf {\bibinfo {volume} {119}},\ \bibinfo
  {pages} {147703} (\bibinfo {year} {2017})}\BibitemShut {NoStop}%
\bibitem [{\citenamefont {Ranzani}\ \emph {et~al.}(2017)\citenamefont
  {Ranzani}, \citenamefont {Kotler}, \citenamefont {Sirois}, \citenamefont
  {DeFeo}, \citenamefont {Castellanos-Beltran}, \citenamefont {Cicak},
  \citenamefont {Vale},\ and\ \citenamefont {Aumentado}}]{ranzani:2017}%
  \BibitemOpen
  \bibfield  {author} {\bibinfo {author} {\bibfnamefont {Leonardo}\
  \bibnamefont {Ranzani}}, \bibinfo {author} {\bibfnamefont {Shlomi}\
  \bibnamefont {Kotler}}, \bibinfo {author} {\bibfnamefont {Adam~J.}\
  \bibnamefont {Sirois}}, \bibinfo {author} {\bibfnamefont {Michael~P.}\
  \bibnamefont {DeFeo}}, \bibinfo {author} {\bibfnamefont {Manuel}\
  \bibnamefont {Castellanos-Beltran}}, \bibinfo {author} {\bibfnamefont
  {Katarina}\ \bibnamefont {Cicak}}, \bibinfo {author} {\bibfnamefont
  {Leila~R.}\ \bibnamefont {Vale}}, \ and\ \bibinfo {author} {\bibfnamefont
  {Jos\'e}\ \bibnamefont {Aumentado}},\ }\bibfield  {title} {\enquote {\bibinfo
  {title} {Wideband isolation by frequency conversion in a josephson-junction
  transmission line},}\ }\href {\doibase 10.1103/PhysRevApplied.8.054035}
  {\bibfield  {journal} {\bibinfo  {journal} {Phys. Rev. Applied}\ }\textbf
  {\bibinfo {volume} {8}},\ \bibinfo {pages} {054035} (\bibinfo {year}
  {2017})}\BibitemShut {NoStop}%
\bibitem [{\citenamefont {Hohenwarter}\ \emph {et~al.}(1993)\citenamefont
  {Hohenwarter}, \citenamefont {Track}, \citenamefont {Drake},\ and\
  \citenamefont {Patt}}]{hohenwarter:1993}%
  \BibitemOpen
  \bibfield  {author} {\bibinfo {author} {\bibfnamefont {G.~K.~G.}\
  \bibnamefont {Hohenwarter}}, \bibinfo {author} {\bibfnamefont {E.~K.}\
  \bibnamefont {Track}}, \bibinfo {author} {\bibfnamefont {R.~E.}\ \bibnamefont
  {Drake}}, \ and\ \bibinfo {author} {\bibfnamefont {R.}~\bibnamefont {Patt}},\
  }\bibfield  {title} {\enquote {\bibinfo {title} {Forty five nanoseconds
  superconducting delay lines},}\ }\href {\doibase 10.1109/77.233499}
  {\bibfield  {journal} {\bibinfo  {journal} {IEEE Transactions on Applied
  Superconductivity}\ }\textbf {\bibinfo {volume} {3}},\ \bibinfo {pages}
  {2804--2807} (\bibinfo {year} {1993})}\BibitemShut {NoStop}%
\bibitem [{\citenamefont {Wang}\ \emph {et~al.}(2005)\citenamefont {Wang},
  \citenamefont {Su}, \citenamefont {Huang},\ and\ \citenamefont
  {Lancaster}}]{wang:2005}%
  \BibitemOpen
  \bibfield  {author} {\bibinfo {author} {\bibfnamefont {Yi}~\bibnamefont
  {Wang}}, \bibinfo {author} {\bibfnamefont {Hieng~Tiong}\ \bibnamefont {Su}},
  \bibinfo {author} {\bibfnamefont {Frederick}\ \bibnamefont {Huang}}, \ and\
  \bibinfo {author} {\bibfnamefont {M.~J.}\ \bibnamefont {Lancaster}},\
  }\bibfield  {title} {\enquote {\bibinfo {title} {Wide-band superconducting
  coplanar delay lines},}\ }\href {\doibase 10.1109/TMTT.2005.850436}
  {\bibfield  {journal} {\bibinfo  {journal} {IEEE Transactions on Microwave
  Theory and Techniques}\ }\textbf {\bibinfo {volume} {53}},\ \bibinfo {pages}
  {2348--2354} (\bibinfo {year} {2005})}\BibitemShut {NoStop}%
\bibitem [{\citenamefont {Su}\ \emph {et~al.}(2008)\citenamefont {Su},
  \citenamefont {Wang}, \citenamefont {Huang},\ and\ \citenamefont
  {Lancaster}}]{su:2008}%
  \BibitemOpen
  \bibfield  {author} {\bibinfo {author} {\bibfnamefont {Hieng~Tiong}\
  \bibnamefont {Su}}, \bibinfo {author} {\bibfnamefont {Yi}~\bibnamefont
  {Wang}}, \bibinfo {author} {\bibfnamefont {Frederick}\ \bibnamefont {Huang}},
  \ and\ \bibinfo {author} {\bibfnamefont {Michael~J.}\ \bibnamefont
  {Lancaster}},\ }\bibfield  {title} {\enquote {\bibinfo {title}
  {Superconducting delay lines},}\ }\href {\doibase 10.1007/s10948-007-0239-2}
  {\bibfield  {journal} {\bibinfo  {journal} {Journal of Superconductivity and
  Novel Magnetism}\ }\textbf {\bibinfo {volume} {21}},\ \bibinfo {pages}
  {7--16} (\bibinfo {year} {2008})}\BibitemShut {NoStop}%
\bibitem [{\citenamefont {Zhong}\ \emph {et~al.}(2018)\citenamefont {Zhong},
  \citenamefont {Chang}, \citenamefont {Satzinger}, \citenamefont {Chou},
  \citenamefont {Bienfait}, \citenamefont {Conner}, \citenamefont {Dumur},
  \citenamefont {Grebel}, \citenamefont {Peairs}, \citenamefont {Povey},
  \citenamefont {Schuster},\ and\ \citenamefont {Cleland}}]{zhong:2018}%
  \BibitemOpen
  \bibfield  {author} {\bibinfo {author} {\bibfnamefont {Y.~P.}\ \bibnamefont
  {Zhong}}, \bibinfo {author} {\bibfnamefont {H.-S.}\ \bibnamefont {Chang}},
  \bibinfo {author} {\bibfnamefont {K.~J.}\ \bibnamefont {Satzinger}}, \bibinfo
  {author} {\bibfnamefont {M.-H.}\ \bibnamefont {Chou}}, \bibinfo {author}
  {\bibfnamefont {A}~\bibnamefont {Bienfait}}, \bibinfo {author} {\bibfnamefont
  {C.~R.}\ \bibnamefont {Conner}}, \bibinfo {author} {\bibfnamefont
  {{\'E}}~\bibnamefont {Dumur}}, \bibinfo {author} {\bibfnamefont
  {J}~\bibnamefont {Grebel}}, \bibinfo {author} {\bibfnamefont {G.~A.}\
  \bibnamefont {Peairs}}, \bibinfo {author} {\bibfnamefont {R.~G.}\
  \bibnamefont {Povey}}, \bibinfo {author} {\bibfnamefont {D.~I.}\ \bibnamefont
  {Schuster}}, \ and\ \bibinfo {author} {\bibfnamefont {A.~N.}\ \bibnamefont
  {Cleland}},\ }\bibfield  {title} {\enquote {\bibinfo {title} {Violating
  {B}ell's inequality with remotely-connected superconducting qubits},}\
  }\href@noop {} {\bibfield  {journal} {\bibinfo  {journal} {arXiv preprint
  arXiv:1808.03000}\ } (\bibinfo {year} {2018})}\BibitemShut {NoStop}%
\bibitem [{\citenamefont {Bode}\ \emph {et~al.}(1945)\citenamefont {Bode} \emph
  {et~al.}}]{bode:1945}%
  \BibitemOpen
  \bibfield  {author} {\bibinfo {author} {\bibfnamefont {Hendrik~Wade}\
  \bibnamefont {Bode}} \emph {et~al.},\ }\href@noop {} {\emph {\bibinfo {title}
  {Network analysis and feedback amplifier design}}}\ (\bibinfo  {publisher}
  {van Nostrand},\ \bibinfo {year} {1945})\BibitemShut {NoStop}%
\bibitem [{\citenamefont {Fano}(1950)}]{fano:1950}%
  \BibitemOpen
  \bibfield  {author} {\bibinfo {author} {\bibfnamefont {Robert~M}\
  \bibnamefont {Fano}},\ }\bibfield  {title} {\enquote {\bibinfo {title}
  {Theoretical limitations on the broadband matching of arbitrary
  impedances},}\ }\href@noop {} {\bibfield  {journal} {\bibinfo  {journal}
  {Journal of the Franklin Institute}\ }\textbf {\bibinfo {volume} {249}},\
  \bibinfo {pages} {57--83,139--154} (\bibinfo {year} {1950})}\BibitemShut
  {NoStop}%
\bibitem [{\citenamefont {Pozar}(2011)}]{pozar:2011}%
  \BibitemOpen
  \bibfield  {author} {\bibinfo {author} {\bibfnamefont {David~M}\ \bibnamefont
  {Pozar}},\ }\href@noop {} {\enquote {\bibinfo {title} {Microwave engineering.
  4th},}\ } (\bibinfo {year} {2011})\BibitemShut {NoStop}%
\bibitem [{\citenamefont {Bell}\ \emph {et~al.}(2012)\citenamefont {Bell},
  \citenamefont {Sadovskyy}, \citenamefont {Ioffe}, \citenamefont {Kitaev},\
  and\ \citenamefont {Gershenson}}]{bell:2012}%
  \BibitemOpen
  \bibfield  {author} {\bibinfo {author} {\bibfnamefont {M.~T.}\ \bibnamefont
  {Bell}}, \bibinfo {author} {\bibfnamefont {I.~A.}\ \bibnamefont {Sadovskyy}},
  \bibinfo {author} {\bibfnamefont {L.~B.}\ \bibnamefont {Ioffe}}, \bibinfo
  {author} {\bibfnamefont {A.~Yu.}\ \bibnamefont {Kitaev}}, \ and\ \bibinfo
  {author} {\bibfnamefont {M.~E.}\ \bibnamefont {Gershenson}},\ }\bibfield
  {title} {\enquote {\bibinfo {title} {Quantum superinductor with tunable
  nonlinearity},}\ }\href {\doibase 10.1103/PhysRevLett.109.137003} {\bibfield
  {journal} {\bibinfo  {journal} {Phys. Rev. Lett.}\ }\textbf {\bibinfo
  {volume} {109}},\ \bibinfo {pages} {137003} (\bibinfo {year}
  {2012})}\BibitemShut {NoStop}%
\bibitem [{\citenamefont {Naaman}\ \emph
  {et~al.}(2017{\natexlab{b}})\citenamefont {Naaman}, \citenamefont {Strong},
  \citenamefont {Ferguson}, \citenamefont {Egan}, \citenamefont {Bailey},\ and\
  \citenamefont {Hinkey}}]{naaman:2016}%
  \BibitemOpen
  \bibfield  {author} {\bibinfo {author} {\bibfnamefont {O.}~\bibnamefont
  {Naaman}}, \bibinfo {author} {\bibfnamefont {J.~A.}\ \bibnamefont {Strong}},
  \bibinfo {author} {\bibfnamefont {D.~G.}\ \bibnamefont {Ferguson}}, \bibinfo
  {author} {\bibfnamefont {J.}~\bibnamefont {Egan}}, \bibinfo {author}
  {\bibfnamefont {N.}~\bibnamefont {Bailey}}, \ and\ \bibinfo {author}
  {\bibfnamefont {R.~T.}\ \bibnamefont {Hinkey}},\ }\bibfield  {title}
  {\enquote {\bibinfo {title} {Josephson junction microwave modulators for
  qubit control},}\ }\href {\doibase 10.1063/1.4976809} {\bibfield  {journal}
  {\bibinfo  {journal} {Journal of Applied Physics}\ }\textbf {\bibinfo
  {volume} {121}},\ \bibinfo {pages} {073904} (\bibinfo {year}
  {2017}{\natexlab{b}})},\ \Eprint
  {http://arxiv.org/abs/https://doi.org/10.1063/1.4976809}
  {https://doi.org/10.1063/1.4976809} \BibitemShut {NoStop}%
\bibitem [{\citenamefont {Kerckhoff}\ \emph {et~al.}(2015)\citenamefont
  {Kerckhoff}, \citenamefont {Lalumi\`ere}, \citenamefont {Chapman},
  \citenamefont {Blais},\ and\ \citenamefont {Lehnert}}]{kerckhoff:2015}%
  \BibitemOpen
  \bibfield  {author} {\bibinfo {author} {\bibfnamefont {Joseph}\ \bibnamefont
  {Kerckhoff}}, \bibinfo {author} {\bibfnamefont {Kevin}\ \bibnamefont
  {Lalumi\`ere}}, \bibinfo {author} {\bibfnamefont {Benjamin~J.}\ \bibnamefont
  {Chapman}}, \bibinfo {author} {\bibfnamefont {Alexandre}\ \bibnamefont
  {Blais}}, \ and\ \bibinfo {author} {\bibfnamefont {K.~W.}\ \bibnamefont
  {Lehnert}},\ }\bibfield  {title} {\enquote {\bibinfo {title} {On-chip
  superconducting microwave circulator from synthetic rotation},}\ }\href
  {\doibase 10.1103/PhysRevApplied.4.034002} {\bibfield  {journal} {\bibinfo
  {journal} {Phys. Rev. Applied}\ }\textbf {\bibinfo {volume} {4}},\ \bibinfo
  {pages} {034002} (\bibinfo {year} {2015})}\BibitemShut {NoStop}%
\bibitem [{\citenamefont {Chapman}\ \emph {et~al.}(2016)\citenamefont
  {Chapman}, \citenamefont {Moores}, \citenamefont {Rosenthal}, \citenamefont
  {Kerckhoff},\ and\ \citenamefont {Lehnert}}]{chapman:2016}%
  \BibitemOpen
  \bibfield  {author} {\bibinfo {author} {\bibfnamefont {Benjamin~J}\
  \bibnamefont {Chapman}}, \bibinfo {author} {\bibfnamefont {Bradley~A}\
  \bibnamefont {Moores}}, \bibinfo {author} {\bibfnamefont {Eric~I}\
  \bibnamefont {Rosenthal}}, \bibinfo {author} {\bibfnamefont {Joseph}\
  \bibnamefont {Kerckhoff}}, \ and\ \bibinfo {author} {\bibfnamefont {K.~W.}\
  \bibnamefont {Lehnert}},\ }\bibfield  {title} {\enquote {\bibinfo {title}
  {General purpose multiplexing device for cryogenic microwave systems},}\
  }\href@noop {} {\bibfield  {journal} {\bibinfo  {journal} {Applied Physics
  Letters}\ }\textbf {\bibinfo {volume} {108}},\ \bibinfo {pages} {222602}
  (\bibinfo {year} {2016})}\BibitemShut {NoStop}%
\bibitem [{\citenamefont {Chapman}\ \emph
  {et~al.}(2017{\natexlab{b}})\citenamefont {Chapman}, \citenamefont
  {Rosenthal}, \citenamefont {Kerckhoff}, \citenamefont {Vale}, \citenamefont
  {Hilton},\ and\ \citenamefont {Lehnert}}]{chapman:2017}%
  \BibitemOpen
  \bibfield  {author} {\bibinfo {author} {\bibfnamefont {Benjamin~J.}\
  \bibnamefont {Chapman}}, \bibinfo {author} {\bibfnamefont {Eric~I.}\
  \bibnamefont {Rosenthal}}, \bibinfo {author} {\bibfnamefont {Joseph}\
  \bibnamefont {Kerckhoff}}, \bibinfo {author} {\bibfnamefont {Leila~R.}\
  \bibnamefont {Vale}}, \bibinfo {author} {\bibfnamefont {Gene~C.}\
  \bibnamefont {Hilton}}, \ and\ \bibinfo {author} {\bibfnamefont {K.~W.}\
  \bibnamefont {Lehnert}},\ }\bibfield  {title} {\enquote {\bibinfo {title}
  {Single-sideband modulator for frequency domain multiplexing of
  superconducting qubit readout},}\ }\href@noop {} {\bibfield  {journal}
  {\bibinfo  {journal} {Applied Physics Letters}\ }\textbf {\bibinfo {volume}
  {110}},\ \bibinfo {pages} {162601} (\bibinfo {year}
  {2017}{\natexlab{b}})}\BibitemShut {NoStop}%
\bibitem [{\citenamefont {Maezawa}\ \emph {et~al.}(1995)\citenamefont
  {Maezawa}, \citenamefont {Aoyagi}, \citenamefont {Nakagawa}, \citenamefont
  {Kurosawa},\ and\ \citenamefont {Takada}}]{maezawa:1995}%
  \BibitemOpen
  \bibfield  {author} {\bibinfo {author} {\bibfnamefont {M.}~\bibnamefont
  {Maezawa}}, \bibinfo {author} {\bibfnamefont {M.}~\bibnamefont {Aoyagi}},
  \bibinfo {author} {\bibfnamefont {H.}~\bibnamefont {Nakagawa}}, \bibinfo
  {author} {\bibfnamefont {I.}~\bibnamefont {Kurosawa}}, \ and\ \bibinfo
  {author} {\bibfnamefont {S.}~\bibnamefont {Takada}},\ }\bibfield  {title}
  {\enquote {\bibinfo {title} {Specific capacitance of {N}b/{A}l{O}x/{N}b
  {J}osephson junctions with critical current densities in the range of
  $0.1-18$ ka/cm$^2$},}\ }\href {\doibase 10.1063/1.113927} {\bibfield
  {journal} {\bibinfo  {journal} {Applied Physics Letters}\ }\textbf {\bibinfo
  {volume} {66}},\ \bibinfo {pages} {2134--2136} (\bibinfo {year} {1995})},\
  \Eprint {http://arxiv.org/abs/https://doi.org/10.1063/1.113927}
  {https://doi.org/10.1063/1.113927} \BibitemShut {NoStop}%
\bibitem [{\citenamefont {Tolpygo}\ \emph {et~al.}(2017)\citenamefont
  {Tolpygo}, \citenamefont {Bolkhovsky}, \citenamefont {Zarr}, \citenamefont
  {Weir}, \citenamefont {Wynn}, \citenamefont {Day}, \citenamefont {Johnson},\
  and\ \citenamefont {Gouker}}]{tolpygo:2017}%
  \BibitemOpen
  \bibfield  {author} {\bibinfo {author} {\bibfnamefont {S.~K.}\ \bibnamefont
  {Tolpygo}}, \bibinfo {author} {\bibfnamefont {V.}~\bibnamefont {Bolkhovsky}},
  \bibinfo {author} {\bibfnamefont {S.}~\bibnamefont {Zarr}}, \bibinfo {author}
  {\bibfnamefont {T.~J.}\ \bibnamefont {Weir}}, \bibinfo {author}
  {\bibfnamefont {A.}~\bibnamefont {Wynn}}, \bibinfo {author} {\bibfnamefont
  {A.~L.}\ \bibnamefont {Day}}, \bibinfo {author} {\bibfnamefont {L.~M.}\
  \bibnamefont {Johnson}}, \ and\ \bibinfo {author} {\bibfnamefont {M.~A.}\
  \bibnamefont {Gouker}},\ }\bibfield  {title} {\enquote {\bibinfo {title}
  {Properties of unshunted and resistively shunted {N}b/{A}l{O}x-{A}l/{N}b
  {J}osephson junctions with critical current densities from 0.1 to 1
  ma/$\mu$m$^2$},}\ }\href {\doibase 10.1109/TASC.2017.2667403} {\bibfield
  {journal} {\bibinfo  {journal} {IEEE Transactions on Applied
  Superconductivity}\ }\textbf {\bibinfo {volume} {27}},\ \bibinfo {pages}
  {1--15} (\bibinfo {year} {2017})}\BibitemShut {NoStop}%
\bibitem [{\citenamefont {Hasnain}\ \emph {et~al.}(1986)\citenamefont
  {Hasnain}, \citenamefont {Dienes},\ and\ \citenamefont
  {Whinnery}}]{hasnain:1986}%
  \BibitemOpen
  \bibfield  {author} {\bibinfo {author} {\bibfnamefont {G.}~\bibnamefont
  {Hasnain}}, \bibinfo {author} {\bibfnamefont {A.}~\bibnamefont {Dienes}}, \
  and\ \bibinfo {author} {\bibfnamefont {J.~R.}\ \bibnamefont {Whinnery}},\
  }\bibfield  {title} {\enquote {\bibinfo {title} {Dispersion of picosecond
  pulses in coplanar transmission lines},}\ }\href {\doibase
  10.1109/TMTT.1986.1133427} {\bibfield  {journal} {\bibinfo  {journal} {IEEE
  Transactions on Microwave Theory and Techniques}\ }\textbf {\bibinfo {volume}
  {34}},\ \bibinfo {pages} {738--741} (\bibinfo {year} {1986})}\BibitemShut
  {NoStop}%
\bibitem [{\citenamefont {Gao}(2008)}]{gao:2008}%
  \BibitemOpen
  \bibfield  {author} {\bibinfo {author} {\bibfnamefont {Jiansong}\
  \bibnamefont {Gao}},\ }\emph {\bibinfo {title} {The physics of
  superconducting microwave resonators}},\ \href@noop {} {Ph.D. thesis},\
  \bibinfo  {school} {California Institute of Technology} (\bibinfo {year}
  {2008})\BibitemShut {NoStop}%
\bibitem [{\citenamefont {Orfanidis}(2002)}]{orfanidis:2002}%
  \BibitemOpen
  \bibfield  {author} {\bibinfo {author} {\bibfnamefont {Sophocles~J}\
  \bibnamefont {Orfanidis}},\ }\enquote {\bibinfo {title} {Electromagnetic
  waves and antennas},}\ \ (\bibinfo  {publisher} {Rutgers University New
  Brunswick, NJ},\ \bibinfo {year} {2002})\BibitemShut {NoStop}%
\bibitem [{\citenamefont {O'Connell}\ \emph {et~al.}(2008)\citenamefont
  {O'Connell}, \citenamefont {Ansmann}, \citenamefont {Bialczak}, \citenamefont
  {Hofheinz}, \citenamefont {Katz}, \citenamefont {Lucero}, \citenamefont
  {McKenney}, \citenamefont {Neeley}, \citenamefont {Wang}, \citenamefont
  {Weig}, \citenamefont {Cleland},\ and\ \citenamefont
  {Martinis}}]{oconnell:2008}%
  \BibitemOpen
  \bibfield  {author} {\bibinfo {author} {\bibfnamefont {A.~D.}\ \bibnamefont
  {O'Connell}}, \bibinfo {author} {\bibfnamefont {M.}~\bibnamefont {Ansmann}},
  \bibinfo {author} {\bibfnamefont {R.~C.}\ \bibnamefont {Bialczak}}, \bibinfo
  {author} {\bibfnamefont {M.}~\bibnamefont {Hofheinz}}, \bibinfo {author}
  {\bibfnamefont {N.}~\bibnamefont {Katz}}, \bibinfo {author} {\bibfnamefont
  {Erik}\ \bibnamefont {Lucero}}, \bibinfo {author} {\bibfnamefont
  {C.}~\bibnamefont {McKenney}}, \bibinfo {author} {\bibfnamefont
  {M.}~\bibnamefont {Neeley}}, \bibinfo {author} {\bibfnamefont
  {H.}~\bibnamefont {Wang}}, \bibinfo {author} {\bibfnamefont {E.~M.}\
  \bibnamefont {Weig}}, \bibinfo {author} {\bibfnamefont {A.~N.}\ \bibnamefont
  {Cleland}}, \ and\ \bibinfo {author} {\bibfnamefont {J.~M.}\ \bibnamefont
  {Martinis}},\ }\bibfield  {title} {\enquote {\bibinfo {title} {Microwave
  dielectric loss at single photon energies and millikelvin temperatures},}\
  }\href {\doibase 10.1063/1.2898887} {\bibfield  {journal} {\bibinfo
  {journal} {Applied Physics Letters}\ }\textbf {\bibinfo {volume} {92}},\
  \bibinfo {pages} {112903} (\bibinfo {year} {2008})},\ \Eprint
  {http://arxiv.org/abs/https://doi.org/10.1063/1.2898887}
  {https://doi.org/10.1063/1.2898887} \BibitemShut {NoStop}%
\bibitem [{\citenamefont {Van~Duzer}\ and\ \citenamefont
  {Turner}(1981)}]{vanduzer:1981}%
  \BibitemOpen
  \bibfield  {author} {\bibinfo {author} {\bibfnamefont {Theodore}\
  \bibnamefont {Van~Duzer}}\ and\ \bibinfo {author} {\bibfnamefont
  {Charles~William}\ \bibnamefont {Turner}},\ }\href@noop {} {\emph {\bibinfo
  {title} {Principles of superconductive devices and circuits}}},\ \bibinfo
  {edition} {2nd}\ ed.\ (\bibinfo  {publisher} {Prentice Hall},\ \bibinfo
  {year} {1981})\BibitemShut {NoStop}%
\bibitem [{\citenamefont {Nagappan}(1970)}]{nagappan:1970}%
  \BibitemOpen
  \bibfield  {author} {\bibinfo {author} {\bibfnamefont {Kasi}\ \bibnamefont
  {Nagappan}},\ }\bibfield  {title} {\enquote {\bibinfo {title} {Step-by-step
  formation of bus admittance matrix},}\ }\href@noop {} {\bibfield  {journal}
  {\bibinfo  {journal} {IEEE transactions on power apparatus and systems}\
  }\textbf {\bibinfo {volume} {89}},\ \bibinfo {pages} {812--820} (\bibinfo
  {year} {1970})}\BibitemShut {NoStop}%
\bibitem [{\citenamefont {Hover}\ \emph {et~al.}(2012)\citenamefont {Hover},
  \citenamefont {Chen}, \citenamefont {Ribeill}, \citenamefont {Zhu},
  \citenamefont {Sendelbach},\ and\ \citenamefont {McDermott}}]{hover:2012}%
  \BibitemOpen
  \bibfield  {author} {\bibinfo {author} {\bibfnamefont {D.}~\bibnamefont
  {Hover}}, \bibinfo {author} {\bibfnamefont {Y.-F.}\ \bibnamefont {Chen}},
  \bibinfo {author} {\bibfnamefont {G.~J.}\ \bibnamefont {Ribeill}}, \bibinfo
  {author} {\bibfnamefont {S.}~\bibnamefont {Zhu}}, \bibinfo {author}
  {\bibfnamefont {S.}~\bibnamefont {Sendelbach}}, \ and\ \bibinfo {author}
  {\bibfnamefont {R.}~\bibnamefont {McDermott}},\ }\bibfield  {title} {\enquote
  {\bibinfo {title} {Superconducting low-inductance undulatory galvanometer
  microwave amplifier},}\ }\href {\doibase 10.1063/1.3682309} {\bibfield
  {journal} {\bibinfo  {journal} {Applied Physics Letters}\ }\textbf {\bibinfo
  {volume} {100}},\ \bibinfo {pages} {063503} (\bibinfo {year} {2012})},\
  \Eprint {http://arxiv.org/abs/https://doi.org/10.1063/1.3682309}
  {https://doi.org/10.1063/1.3682309} \BibitemShut {NoStop}%
\bibitem [{\citenamefont {Caves}(1982)}]{caves:1982}%
  \BibitemOpen
  \bibfield  {author} {\bibinfo {author} {\bibfnamefont {Carlton~M.}\
  \bibnamefont {Caves}},\ }\bibfield  {title} {\enquote {\bibinfo {title}
  {Quantum limits on noise in linear amplifiers},}\ }\href {\doibase
  10.1103/PhysRevD.26.1817} {\bibfield  {journal} {\bibinfo  {journal} {Phys.
  Rev. D}\ }\textbf {\bibinfo {volume} {26}},\ \bibinfo {pages} {1817--1839}
  (\bibinfo {year} {1982})}\BibitemShut {NoStop}%
\bibitem [{\citenamefont {Yurke}\ and\ \citenamefont
  {Denker}(1984)}]{yurke:1984}%
  \BibitemOpen
  \bibfield  {author} {\bibinfo {author} {\bibfnamefont {Bernard}\ \bibnamefont
  {Yurke}}\ and\ \bibinfo {author} {\bibfnamefont {John~S}\ \bibnamefont
  {Denker}},\ }\bibfield  {title} {\enquote {\bibinfo {title} {Quantum network
  theory},}\ }\href@noop {} {\bibfield  {journal} {\bibinfo  {journal}
  {Physical Review A}\ }\textbf {\bibinfo {volume} {29}},\ \bibinfo {pages}
  {1419} (\bibinfo {year} {1984})}\BibitemShut {NoStop}%
\bibitem [{\citenamefont {Lanczos}(1988)}]{lanczos:1988}%
  \BibitemOpen
  \bibfield  {author} {\bibinfo {author} {\bibfnamefont {Cornelius}\
  \bibnamefont {Lanczos}},\ }\enquote {\bibinfo {title} {Applied analysis},}\ \
  (\bibinfo  {publisher} {Courier Corporation},\ \bibinfo {year} {1988})\ pp.\
  \bibinfo {pages} {225--229}\BibitemShut {NoStop}%
\bibitem [{\citenamefont {Yeh}\ \emph {et~al.}(2017)\citenamefont {Yeh},
  \citenamefont {LeFebvre}, \citenamefont {Premaratne}, \citenamefont
  {Wellstood},\ and\ \citenamefont {Palmer}}]{yeh:2017}%
  \BibitemOpen
  \bibfield  {author} {\bibinfo {author} {\bibfnamefont {Jen-Hao}\ \bibnamefont
  {Yeh}}, \bibinfo {author} {\bibfnamefont {Jay}\ \bibnamefont {LeFebvre}},
  \bibinfo {author} {\bibfnamefont {Shavindra}\ \bibnamefont {Premaratne}},
  \bibinfo {author} {\bibfnamefont {F.~C.}\ \bibnamefont {Wellstood}}, \ and\
  \bibinfo {author} {\bibfnamefont {B.~S.}\ \bibnamefont {Palmer}},\ }\bibfield
   {title} {\enquote {\bibinfo {title} {Microwave attenuators for use with
  quantum devices below 100 m{K}},}\ }\href {\doibase 10.1063/1.4984894}
  {\bibfield  {journal} {\bibinfo  {journal} {Journal of Applied Physics}\
  }\textbf {\bibinfo {volume} {121}},\ \bibinfo {pages} {224501} (\bibinfo
  {year} {2017})},\ \Eprint
  {http://arxiv.org/abs/https://doi.org/10.1063/1.4984894}
  {https://doi.org/10.1063/1.4984894} \BibitemShut {NoStop}%
\bibitem [{\citenamefont {Yan}\ \emph {et~al.}(2018)\citenamefont {Yan},
  \citenamefont {Campbell}, \citenamefont {Krantz}, \citenamefont {Kjaergaard},
  \citenamefont {Kim}, \citenamefont {Yoder}, \citenamefont {Hover},
  \citenamefont {Sears}, \citenamefont {Kerman}, \citenamefont {Orlando},
  \citenamefont {Gustavsson},\ and\ \citenamefont {Oliver}}]{yan:2018}%
  \BibitemOpen
  \bibfield  {author} {\bibinfo {author} {\bibfnamefont {Fei}\ \bibnamefont
  {Yan}}, \bibinfo {author} {\bibfnamefont {Dan}\ \bibnamefont {Campbell}},
  \bibinfo {author} {\bibfnamefont {Philip}\ \bibnamefont {Krantz}}, \bibinfo
  {author} {\bibfnamefont {Morten}\ \bibnamefont {Kjaergaard}}, \bibinfo
  {author} {\bibfnamefont {David}\ \bibnamefont {Kim}}, \bibinfo {author}
  {\bibfnamefont {Jonilyn~L.}\ \bibnamefont {Yoder}}, \bibinfo {author}
  {\bibfnamefont {David}\ \bibnamefont {Hover}}, \bibinfo {author}
  {\bibfnamefont {Adam}\ \bibnamefont {Sears}}, \bibinfo {author}
  {\bibfnamefont {Andrew~J.}\ \bibnamefont {Kerman}}, \bibinfo {author}
  {\bibfnamefont {Terry~P.}\ \bibnamefont {Orlando}}, \bibinfo {author}
  {\bibfnamefont {Simon}\ \bibnamefont {Gustavsson}}, \ and\ \bibinfo {author}
  {\bibfnamefont {William~D.}\ \bibnamefont {Oliver}},\ }\bibfield  {title}
  {\enquote {\bibinfo {title} {Distinguishing coherent and thermal photon noise
  in a circuit quantum electrodynamical system},}\ }\href {\doibase
  10.1103/PhysRevLett.120.260504} {\bibfield  {journal} {\bibinfo  {journal}
  {Phys. Rev. Lett.}\ }\textbf {\bibinfo {volume} {120}},\ \bibinfo {pages}
  {260504} (\bibinfo {year} {2018})}\BibitemShut {NoStop}%
\bibitem [{\citenamefont {Wang}\ \emph {et~al.}(2018)\citenamefont {Wang},
  \citenamefont {Shankar}, \citenamefont {Minev}, \citenamefont {P.},
  \citenamefont {A.},\ and\ \citenamefont {Devoret}}]{wang:2018}%
  \BibitemOpen
  \bibfield  {author} {\bibinfo {author} {\bibfnamefont {Z.}~\bibnamefont
  {Wang}}, \bibinfo {author} {\bibfnamefont {S.}~\bibnamefont {Shankar}},
  \bibinfo {author} {\bibfnamefont {Z.~K.}\ \bibnamefont {Minev}}, \bibinfo
  {author} {\bibfnamefont {Campagne-Ibarcq}\ \bibnamefont {P.}}, \bibinfo
  {author} {\bibfnamefont {Narla}\ \bibnamefont {A.}}, \ and\ \bibinfo {author}
  {\bibfnamefont {Michel~H}\ \bibnamefont {Devoret}},\ }\bibfield  {title}
  {\enquote {\bibinfo {title} {Cavity attenuators for superconducting
  qubits},}\ }\href@noop {} {\bibfield  {journal} {\bibinfo  {journal} {arXiv
  preprint arXiv:1807.04849}\ } (\bibinfo {year} {2018})}\BibitemShut {NoStop}%
\end{thebibliography}
\end{document}